\documentclass[12pt]{cernprep}
\usepackage{graphicx}
\usepackage{amssymb}
\usepackage{epsfig,color,rotating}
\usepackage{epstopdf}
\begin{document}
\newcommand{\pT}{\mbox{$p_{\rm T}$}}
\newcommand{\GeVc}{\mbox{GeV/{\it c}}}
\newcommand{\MeVc}{\mbox{MeV/{\it c}}}
\newcommand{\gam}{\mbox{$\gamma$}}
\newcommand{\Deltar}{\mbox{$\Delta r$}}
\newcommand{\Deltat}{\mbox{$\Delta t$}}
\newcommand{\nue}{\mbox{$\nu_{\rm e}$}}
\newcommand{\num}{\mbox{$\nu_{\mu}$}}
\newcommand{\anue}{\mbox{$\bar{\nu}_{\rm e}$}}
\newcommand{\anum}{\mbox{$\bar{\nu}_{\mu}$}}
\newcommand{\anuetoanum}{\mbox{$\bar{\nu}_{\rm e} \rightarrow \bar{\nu}_{\mu}$}}
\newcommand{\numtonue}{\mbox{$\nu_{\mu} \rightarrow \nu_{\rm e}$}}
\newcommand{\anumtoanue}{\mbox{$\bar{\nu}_{\mu} \rightarrow \bar{\nu}_{\rm e}$}}
\newcommand{\pip}{\mbox{$\pi^+$}}
\newcommand{\pim}{\mbox{$\pi^-$}}
\newcommand{\mup}{\mbox{$\mu^+$}}
\newcommand{\mum}{\mbox{$\mu^-$}}
\newcommand{\eplus}{\mbox{e$^+$}}
\newcommand{\eminus}{\mbox{e$^-$}}
\newcommand{\Cdouze}{\mbox{$^{12}$C}}
\newcommand{\Ndouze}{\mbox{$^{12}$N}}
\newcommand{\Bdouze}{\mbox{$^{12}$B}}
\newcommand{\Rgam}{\mbox{$R_{\gamma}$}}
\newcommand{\Rbeta}{\mbox{$R_{\beta}$}}
\newcommand{\signalreaction}{\mbox{$\bar{\nu}_{\rm e}$ + p $\rightarrow$ e$^+$ + n}}
\begin{titlepage}
\docnum{CERN--PH--EP--2011--200}
\date{17 October 2011}
\begin{flushright}
\end{flushright} 
\vspace{1cm}
\title{\large{Revisiting the `LSND anomaly' II: \\ 
critique of the data analysis}}

\begin{abstract}
This paper, together with a preceding paper, questions the so-called `LSND anomaly': a 3.8~$\sigma$
excess of $\bar{\nu}_{\rm e}$ interactions over standard backgrounds, observed by the LSND Collaboration in a beam dump experiment with 800~MeV protons.
That excess has been interpreted as evidence 
for the \anumtoanue\ oscillation in the $\Delta m^2$ range from 0.2~eV$^2$ to 2~eV$^2$. Such a $\Delta m^2$ range is incompatible with the widely accepted model of oscillations between three light neutrino species and would  require the existence of at least one light `sterile' neutrino. In a preceding paper,  it was concluded that the estimates of standard backgrounds must be significantly increased. In this paper, the LSND Collaboration's estimate of the number of $\bar{\nu}_{\rm e}$ interactions followed by neutron capture, and of its error, is questioned. The overall conclusion is that the significance of the `LSND anomaly' is not larger than 2.3~$\sigma$.
\end{abstract}

\vfill  \normalsize
\begin{center}
The HARP--CDP group  \\  

\vspace*{2mm} 

A.~Bolshakova$^1$, 
I.~Boyko$^1$, 
G.~Chelkov$^{1a}$, 
D.~Dedovitch$^1$, 
A.~Elagin$^{1b}$, 
D.~Emelyanov$^1$,
M.~Gostkin$^1$,
A.~Guskov$^1$, 
Z.~Kroumchtein$^1$, 
Yu.~Nefedov$^1$, 
K.~Nikolaev$^1$, 
A.~Zhemchugov$^1$, 
F.~Dydak$^2$, 
J.~Wotschack$^{2*}$, 
A.~De~Min$^{3c}$,
V.~Ammosov$^{4\dagger}$, 
V.~Gapienko$^4$, 
V.~Koreshev$^4$, 
A.~Semak$^4$, 
Yu.~Sviridov$^4$, 
E.~Usenko$^{4d}$, 
V.~Zaets$^4$ 
\\
 
\vspace*{8mm} 

$^1$~{\bf Joint Institute for Nuclear Research, Dubna, Russia} \\
$^2$~{\bf CERN, Geneva, Switzerland} \\ 
$^3$~{\bf Politecnico di Milano and INFN, 
Sezione di Milano-Bicocca, Milan, Italy} \\
$^4$~{\bf Institute of High Energy Physics, Protvino, Russia} \\

\vspace*{5mm}

\submitted{(To be submitted to Phys. Rev. D)}
\end{center}

\vspace*{5mm}
\rule{0.9\textwidth}{0.2mm}

\begin{footnotesize}

$^a$~Also at the Moscow Institute of Physics and Technology, Moscow, Russia 

$^b$~Now at Texas A\&M University, College Station, USA 

$^c$~On leave of absence

$^d$~Now at Institute for Nuclear Research RAS, Moscow, Russia

$^{\dagger}$~Deceased

$^*$~Corresponding author; e-mail: joerg.wotschack@cern.ch
\end{footnotesize}

\end{titlepage}


\newpage 

\section{Introduction}

This is the second of two papers that argue that the 3.8~$\sigma$ significance of the `LSND anomaly', claimed by the LSND Collaboration, cannot be upheld. 

The LSND experiment was carried out at the Los Alamos National Laboratory in the years 1993--1998.
Its scientific goal was a search for \anumtoanue\ oscillations in the `appearance' mode. The neutrino
fluxes were produced by dumping 800~MeV protons into a `beam stop'.  While \num , \anum\ and
\nue\  fluxes were abundant, the \anue\ flux was vanishingly small.

The LSND Collaboration claimed an excess of \anue\ interactions 
over the expectation from standard backgrounds~\cite{LSNDPRD64}. This excess was interpreted as evidence for 
the \anumtoanue\ oscillation with $\Delta m^2$ in the range from 0.2~eV$^2$ to 2~eV$^2$ and came to be known as `LSND anomaly'.
In stark conflict with the widely accepted model of oscillations of three light neutrino flavours,
the excess would require the existence of at least one light `sterile' neutrino that does not couple to
the Z boson.

Since the `LSND anomaly' calls the Standard Model of particle physics in a non-trivial way into question,
the MiniBooNE experiment at the Fermi National Accelerator Laboratory set out to check this result.
In the neutrino mode, an oscillation \numtonue\ with parameters compatible with the LSND claim was 
not seen~\cite{MiniBooNEneutrino}. However, first results from running in antineutrino mode and
searching for \anumtoanue\ led the
MiniBooNE Collaboration to conclude that their result does not rule out 
the `LSND anomaly'~\cite{MiniBooNEantineutrino} that had
indeed been observed in the antineutrino mode.

In this situation it appears worthwhile to undertake a critical review of the original results of the 
LSND experiment that gave rise to the `LSND anomaly'.
This is the subject both of this paper and of a preceding paper~\cite{firstpaper}.

Although we agree with many of the LSND Collaboration's approaches and results, 
in two areas we disagree with LSND. The first area---which 
has been discussed in Ref.~\cite{firstpaper}---concerns the underestimation of the \anue\ flux from
standard sources. According to our assessment the increased standard background reduces the significance of the `LSND anomaly' from 3.8 to 2.9~$\sigma$. The second area is quantitatively discussed in this paper. It concerns the neglect of a bias and the underestimation of systematic errors in the isolation of the signal of $\sim$120 \signalreaction\ reactions with a correlated \gam\  from neutron capture out of $\sim$2100 candidate events.

\section{The LSND fit of the $R_{\gamma}$ distribution}

In this section, we recall LSND's determination of the number of beam-related \anue\ interactions by a fit to the $R_{\gamma}$ distribution---the centre piece of the LSND data analysis. Before, we describe briefly the salient features of the LSND experiment.

The LSND neutrino detector~\cite{LSNDdetector1997}  is a tank filled with liquid scintillator. The amplitude and the time of arrival of scintillation light and of Cherenkov light are detected by an array of phototubes (PMTs)  on the surfaces of the tank. The reconstruction of the energy and position of reaction secondaries is solely based on the responses of the phototubes. There is no distinction between positive and negative electric charges.

The \anue\ flux is measured through the reaction \signalreaction\ which has a well known cross-section. The signature is an \eplus\ and a delayed 2.2~MeV \gam\ from deuteron creation by the capture of the neutron on a free proton, $n$ + $p$ $\rightarrow$ $d$ + $\gamma$. The average delay of the \gam\ is $\sim$186~$\mu$s, determined by the neutron capture cross-section in the scintillator medium.

Besides cuts that enrich beam-induced electrons\footnote{We use the generic term `electron' both for electrons and positrons.} w.r.t. cosmic-ray particles, and a cut that requires electrons to be inside a fiducial volume that ends 35~cm from the faces of the PMTs, a cut 20~MeV $< E_{\rm e} <$ 60~MeV\footnote{The maximum energy of a \anue\ in the decay of a muon at rest is 52.8~MeV.} is applied to qualify as `primary electron'. Furthermore, activities within 1~ms after the primary electron are recorded with a lower trigger threshold not to miss 2.2 MeV $\gamma$'s. The number of such activities can be zero, one, or larger than one. 

Despite the cuts to suppress cosmic-ray particles, there remains a large cosmogenic background. This background is determined in beam-off gates and subtracted after proper normalization from the event numbers observed in beam-on gates. This doubles approximately the statistical error but introduces no systematic uncertainty.

Integrated over the data taking of the LSND experiment in the years 1993 -- 1998, the number of beam-related `primary electrons' is about 2100. The key issue is how to isolate from these 2100 events the signal, i.e., the number of events with a correlated 2.2~MeV \gam\ from neutron capture. We recall that the signal claimed by LSND is 
$117.9$ events~\cite{LSNDPRD64}.

The isolation criterion applied by LSND is the following. ``Correlated 2.2~MeV \gam\ from neutron capture are distinguished from accidental \gam\ from radioactivity by use of the likelihood ratio, \Rgam , which is defined to be the likelihood that the \gam\ is correlated divided by the likelihood that the \gam\ is accidental. \Rgam\ depends on three quantities:  the number of hit PMTs associated with the \gam\ (the multiplicity is proportional to the \gam\ energy), the distance between the reconstructed \gam\ position and positron position, and the time interval between the \gam\ and positron (neutrons have a capture time in mineral oil of 186~$\mu$s, while the accidental \gam\ are 
uniform in time)'' (Section VII.C in Ref.~\cite{LSNDPRD64}).

\begin{figure*}
\begin{center}
\includegraphics[width=0.8\textwidth]{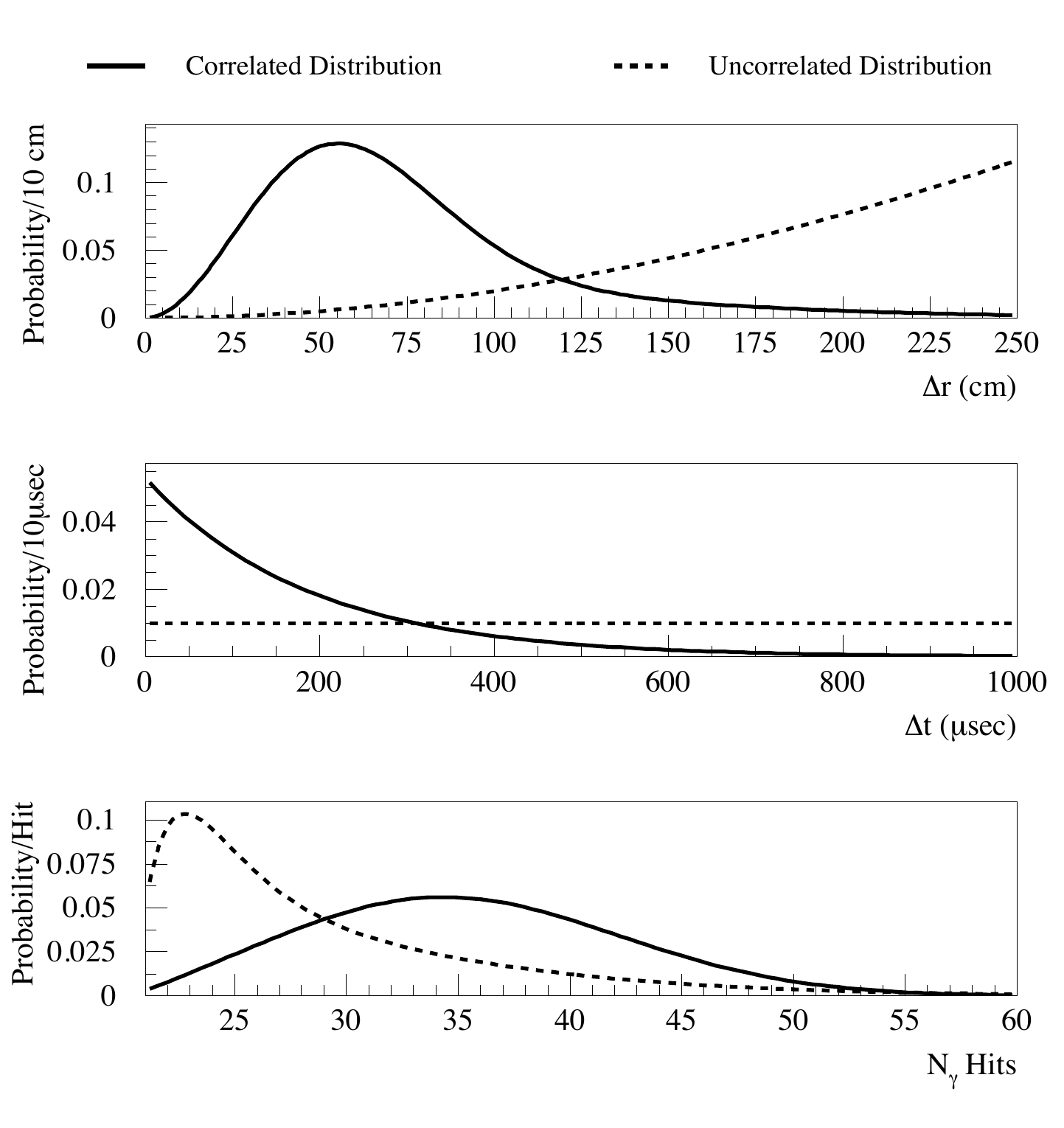}    
\caption{``Distributions for correlated 2.2~MeV $\gamma$ (solid curves) and accidental $\gamma$ (dashed curves). The top plot shows the distance between the reconstructed $\gamma$ position and positron position, $\Delta r$, the middle plot shows the time interval between the $\gamma$ and positron, $\Delta t$, and the bottom plot shows the number of hit phototubes associated with the $\gamma$, $N_{\rm hits}$.'' (Figure and its caption copied from Fig.~10 in Ref.~\cite{LSNDPRD64}.)}.
\label{Fig10inPRD64}
\end{center}
\end{figure*}

\begin{figure*}
\begin{center}
\includegraphics[width=0.8\textwidth]{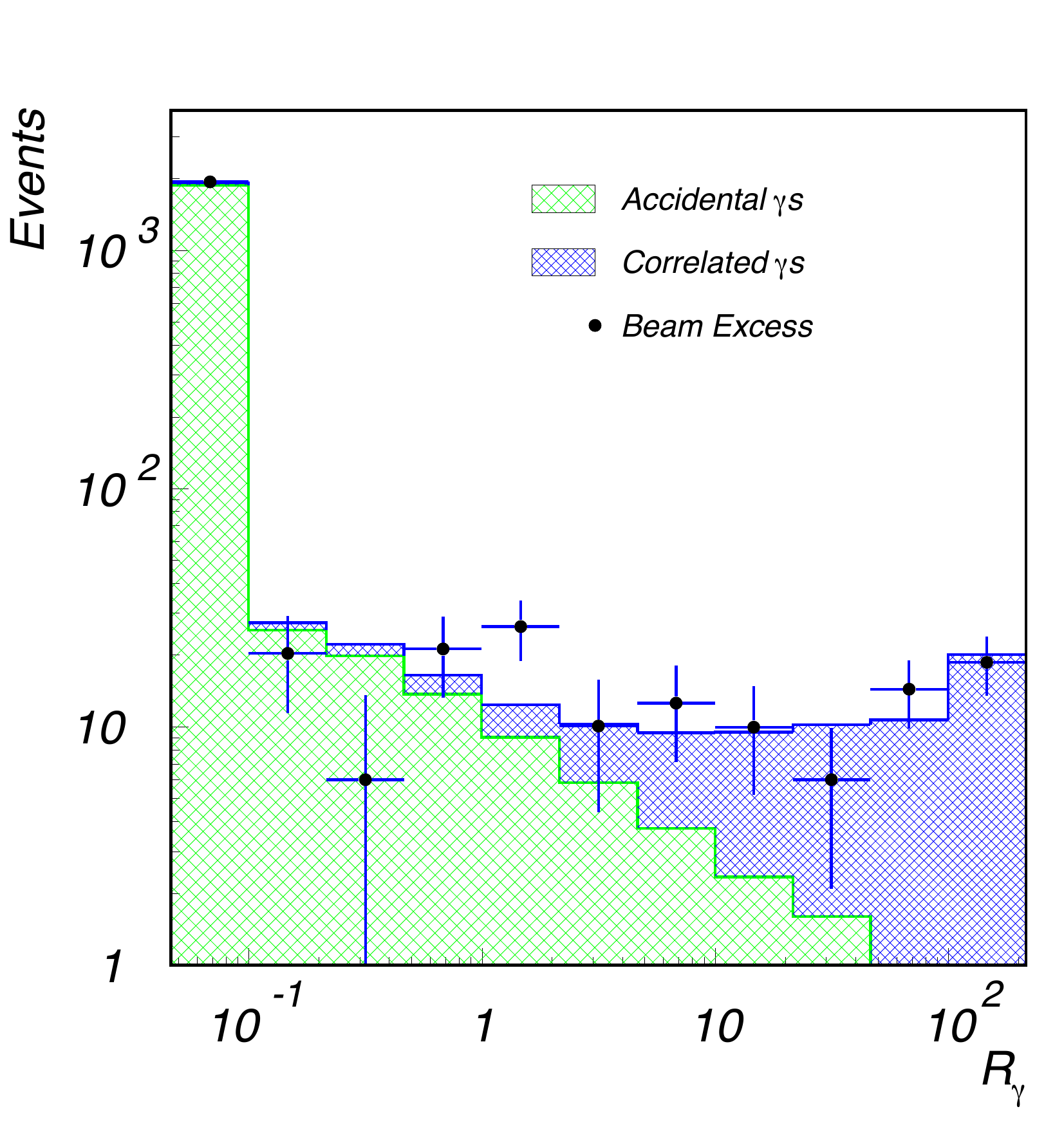}    
\caption{``The \Rgam\ distribution for events that satisfy the selection criteria for the primary \anumtoanue\ oscillation search.'' (Figure and its caption copied from Fig.~14 in Ref.~\cite{LSNDPRD64}.)}.
\label{Fig14inPRD64}
\end{center}
\end{figure*}

The said likelihoods for correlated and accidental $\gamma$'s are obtained from respective distributions of hit PMTs,
distance and time interval that are shown in Fig.~10 of Ref.~\cite{LSNDPRD64} and reproduced here in 
Fig.~\ref{Fig10inPRD64}. In the following, we refer to these distributions as `base distributions'.
An activity within 1~ms after the primary electron is only accepted if three criteria are met: $\Delta r \leq 250$~cm, 8~$\mu$s $ \leq \Delta t \leq $ 1~ms, and $21 \leq N_{\rm hits} \leq 60$. If all three criteria are met, the product of the probabilities of the observed $\Delta r$, $\Delta t$ and $N_{\rm hits}$ is calculated both for the correlated and the accidental distributions, and then their ratio \Rgam . If at least one of the three criteria is not met, the likelihood ratio \Rgam\ is artificially set to zero\footnote{LSND make no statement about lower and upper limits of \Rgam ; in our analysis values of \Rgam\ below 0.051 are set to 0.051, and values above 199 are set to 199.}. If there are several accepted activities, the
activity with the largest \Rgam\ is chosen (Section III.F in Ref.~\cite{LSNDPRC54}). The resulting \Rgam\ distribution of the data is shown as black dots with (statistical) error bars in Fig.~14 in Ref.~\cite{LSNDPRD64}, reproduced here in 
Fig.~\ref{Fig14inPRD64}. 

The prominent peak of the \Rgam\ distribution in the first bin reflects the many events with no accepted activity 
within 1~ms after the primary electron.

The \Rgam\ distribution of the data is then fitted with a linear combination of two \Rgam\ hypotheses: the 
normalized \Rgam\ distribution of correlated $\gamma$'s and the normalized \Rgam\ distribution of accidental $\gamma$'s. The former is obtained by generating triplets [$\Delta r$, $\Delta t$, $N_{\rm hits}$] according to the respective distributions for correlated $\gamma$'s,
and then calculating \Rgam . The latter is analogously obtained by generating triplets [$\Delta r$, $\Delta t$, 
$N_{\rm hits}$] according to the respective distributions for accidental $\gamma$'s, and then calculating \Rgam .
By construction, the \Rgam\ distribution of correlated $\gamma$'s favours large values of \Rgam\ while the \Rgam\ distribution of accidental $\gamma$'s favours small values of \Rgam .  In addition, for either hypothesis an estimate was made of the probability of meeting all three criteria on $\Delta r$, $\Delta t$ and $N_{\rm hits}$ (termed `efficiency'
by LSND).

Then, according to LSND, a $\chi^2$ minimization determines a signal (termed `beam excess' by LSND) of $117.9 \pm 22.4$ events with a correlated \gam\ out of about 2100 candidate events (Section VII.D in Ref.~\cite{LSNDPRD64}).

This procedure, while {\em a priori\/} adequate, is criticized below on several accounts. The results of the procedure depend on the correctness of the `base distributions' 
and the `efficiencies' of correlated and accidental $\gamma$'s, both not discussed in the LSND papers. 
A contribution of positrons from $^{12}$N$_{\rm gs}$ beta decays that are interpreted as correlated $\gamma$'s, has been neglected.

\section{Emulation of the LSND fit of the $R_{\gamma}$ distribution}

With a view to gaining insight into the salient features of the LSND fit of the \Rgam\ distribution by an emulation, 
Table~\ref{DigitizationofFig14inPRD64} gives the numerical values of the bin contents in the double-logarithmic Fig.~\ref{Fig14inPRD64} as obtained by a plot digitization program.
\begin{table*}[htdp]
\caption{Numerical values of the bin contents  in Fig.~\ref{Fig14inPRD64} for the $R_{\gamma}$ distributions of correlated $\gamma$'s, accidental $\gamma$'s, and of the data including errors.}
\begin{center}
\begin{tabular}{|c|c|c|c|}
\hline
$R_{\gamma}$ bin  & Correlated $\gamma$'s & Accidental $\gamma$'s & Data  \\
\hline
\hline
0.050 -- 0.106         &  (50)   &  (1902) & $ (1932 \pm  79)$   \\
0.106 -- 0.226         &   1.70   &  25.48   &  $20.54      \pm    8.65$   \\
0.226 -- 0.480          &  1.81    & 20.10  &   $6.00       \pm   7.63$   \\
0.480 -- 1.02         &   2.57  & 13.63   &  $ 20.98      \pm    7.84$   \\
1.02 -- 2.17          & 3.19   &  9.04     &   $26.04       \pm   7.45$   \\
2.17 -- 4.61           &  4.20    &  5.87  & $10.07       \pm   5.54$  \\
4.61 -- 9.80           &  5.71   &  3.73   &  $12.50       \pm   5.45$   \\
9.80 -- 20.8           &  7.49     &  2.37 &  $10.07        \pm  4.85$   \\
20.8 -- 44.3       &  8.46     & 1.61     &   $6.00       \pm   3.89$   \\
44.3 -- 94.1       &   9.72     &  0.80    &  $14.23       \pm   4.49$   \\
94.1 -- 200       &  19.70    &  0.40   &  $ 18.43       \pm   5.13$   \\
\hline
\end{tabular}
\end{center}
\label{DigitizationofFig14inPRD64}
\end{table*}

For the size of the bin contents and the logarithmic scale, the contents of the first \Rgam\ bin (and, if applicable, their errors) cannot be read off precisely. The values used in our fit emulation are shown in brackets in Table~\ref{DigitizationofFig14inPRD64}.
Thereby, for the two hypothesis distributions use was made of the statement that the correlated and accidental `efficiencies' for
$R_{\gamma} > 1$ are quoted by LSND as 0.51 and 0.012, respectively (Table IX in Ref.~\cite{LSNDPRD64}). 

\begin{figure*}
\begin{center}
\includegraphics[width=0.8\textwidth]{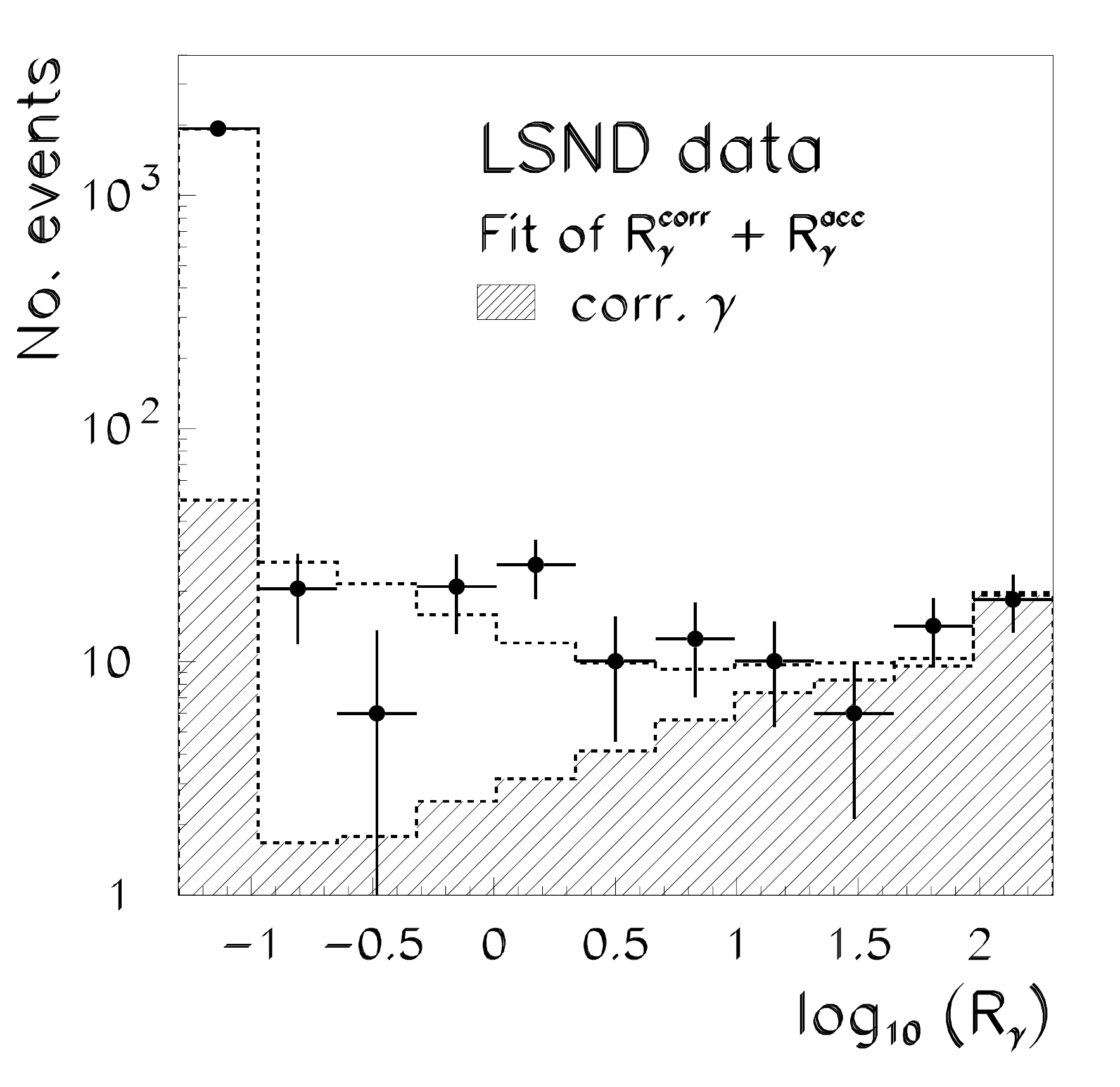} 
\caption{Emulation of the fit of the LSND $R_{\gamma}$ distribution with the sum of the hypotheses for correlated (hatched) and accidental (open) $\gamma$'s.}
\label{11binfitofLSNDdata}
\end{center}
\end{figure*}

The results for the `beam excess' from our fits of the data points shown in Fig.~\ref{Fig14inPRD64} are listed in 
Table~\ref{FitsofFig14data}. Fit No.~2, which is our emulation of the LSND fit, gives the result $112.5 \pm 21.6$ events which is to compared with the published LSND result, listed as Fit No.~1, of $117.9 \pm 22.4$ events. In view of the numerical uncertainties of the contents of the first \Rgam\ bins, we consider our Fit No.~2 a satisfactory emulation of Fit No.~1\footnote{The (unreadable) error of the first $R_{\gamma}$ bin does not matter much: the fit result of $112.5 \pm 21.6$ events changes to $112.4 \pm 21.5$ for an error equal to the square root of the bin content, and
to $112.8 \pm 21.6$ for an error three times the square root of the bin content.}. Fit No.~2 is graphically shown in 
Fig.~\ref{11binfitofLSNDdata}.

The first important aspect of our emulation is that it nearly reproduces the total error of the `beam excess'
(21.6 w.r.t. to 22.4 published by LSND) by taking into account merely the statistical errors of the data points shown in 
Fig.~\ref{Fig14inPRD64}. We recall that the error of 22.4 events was used by LSND to calculate the significance of the `LSND anomaly'. We note that in comparison to the statistical error a vanishingly small systematic error has been taken into account by LSND, if any. 
\begin{table*}[htdp]
\caption{Fits of LSND data shown in Fig.~\ref{Fig14inPRD64}}
\begin{center}
\begin{tabular}{|c|c|c|c|c|}
\hline
Fit No.   &  No. of events with corr. $\gamma$  & $\chi^2$ & d.o.f. & Comment \\
 \hline
\hline
1 & $117.9 \pm 22.4$ (19.0\%) & 10.7 & 9 & Result published by LSND in Ref.~\cite{LSNDPRD64} \\
2 & $112.5 \pm 21.6$ (19.2\%) & 10.8 & 9 & Our emulation of the LSND fit \\
2a & $116.7 \pm 22.1$ (18.9\%) & 10.1 & 8 & Fit No.~2 without 1st $R_{\gamma}$ bin \\
3 & $103.1 \pm 20.2$ (19.6\%) & 11.9 & 9 & $R_{\gamma}$ hypotheses from Fig.~10 in Ref.~\cite{LSNDPRC54} \\
3a & $105.7 \pm 20.5$ (19.4\%) & 11.4 & 8 & Fit No.~3 without 1st $R_{\gamma}$ bin  \\
\hline
\end{tabular}
\end{center}
\label{FitsofFig14data}
\end{table*}

The prominent spikes in the first \Rgam\ bin of Fig.~\ref{Fig14inPRD64}, caused by the `efficiencies' of correlated and uncorrelated $\gamma$'s, raise questions on respective uncertainties and their effect on the `beam excess' and its error.  

Following LSND's nomenclature, the `efficiency' of correlated and uncorrelated $\gamma$'s is defined as the
probability to find a respective $\gamma$ that meets three criteria:  distance $\Delta r \leq$ 250~cm, 
time delay 8 $\mu$s $\leq \Delta t \leq$ 1~ms with respect to the primary electron, and 
pulseheight 20 $\leq N_{\rm hits}  \leq$ 60. The numerical values of these efficiencies are, for $R_{\gamma} > 1$,
0.51 for correlated $\gamma$'s and 0.012 for accidental $\gamma$'s, respectively. Both values are quoted by LSND  with an error of 7\% (Table IX in Ref.~\cite{LSNDPRD64}). 

First, we show that especially the numerical value of the `efficiency' of correlated $\gamma$'s is important for the `beam excess' and its error.

The content of the first bin consists almost exclusively of events that have no accepted delayed \gam . Therefore, the content of the first bin does not contribute to the fit's purpose of discriminating between correlated and accidental $\gamma$'s. So why did LSND include the first $R_{\gamma}$ bin into the fit?
Omitting the first $R_{\gamma}$ bin reduces the number of data bins from 11 to 10, and reduces the degrees of freedom from 9 to 8, but preserves fully the potential of discriminating between correlated and uncorrelated $\gamma$'s. The result for the `beam excess' is
$116.7 \pm 22.1$ events (Fit No.~3 in Table~\ref{FitsofFig14data}) where an `efficiency' of correlated $\gamma$'s of 0.51 for $R_{\gamma} > 1$ has been used.
This result which bypasses the reading uncertainty of the data content of the first bin, is remarkably close to
LSND's result for the bin excess from their 11-bin fit, $117.9 \pm 22.4$ (Fit No.~1 in Table~\ref{FitsofFig14data}), and proves that our emulation of the LSND fit makes sense.

\begin{figure*}
\begin{center}
\includegraphics[width=0.8\textwidth]{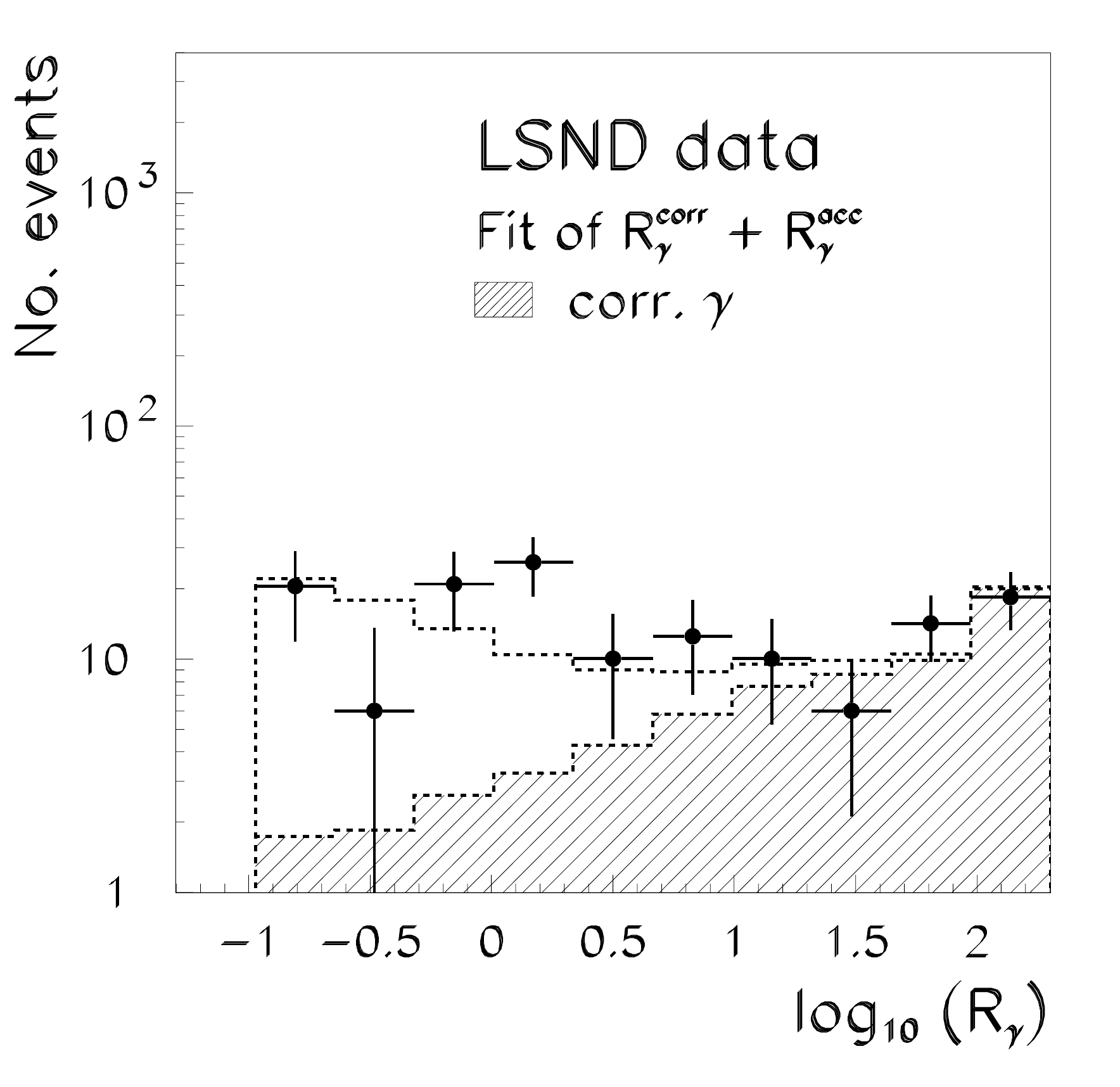}     
\caption{10-bin fit of the LSND $R_{\gamma}$ distribution with the sum of the hypotheses for correlated (hatched) and accidental (open) $\gamma$'s.}
\label{10binfitofLSNDdata}
\end{center}
\end{figure*}

Fit No.~3 is graphically shown in Fig.~\ref{10binfitofLSNDdata}.

Our 10-bin fit sharpens the argument that the LSND fit result comprises little systematic error margin, if at all.
Already the error of 7\% quoted by LSND for the `efficiency' of correlated $\gamma$'s (which will be disputed below)
would increase, quadratically added,  the 22.1 error of our 10-bin fit to 23.6 which alone (other systematic errors still ignored) exceeds already the total error of 22.4 quoted by LSND.

As far as the statistical error is concerned, there is no appreciable gain from the inclusion into the fit of the data in the first $R_{\gamma}$ bin.
As for the systematic error, while the 7\% uncertainty of the `efficiency' of correlated $\gamma$'s cannot be avoided, the propagation of the 7\% systematic uncertainty on the `efficiency' of accidental $\gamma$'s into the error of the `beam excess' could be avoided.

While we consider in retrospect a 10-bin fit as the better choice, we limit ourselves in the further discussion to the 
11-bin fit since this type of fit was LSND's choice and underlies the `LSND anomaly'. 

The `efficiencies' are in an 11-bin fit an integral part of the two hypothesis distributions for correlated and uncorrelated $\gamma$'s. The numerical values of the `efficiencies', in particular the one of correlated $\gamma$'s, have a direct impact on the `beam excess' and its error. 

Where do the `efficiency' values and their errors come from? Can the values quoted by LSND be understood?
It is surprising that the values are quoted in LSND's final physics paper (Table IX in Ref.~\cite{LSNDPRD64}), however, nowhere in this paper is their origin discussed.
The only hint can be found in an earlier paper that quotes intermediate results (Section III.F in Ref.~\cite{LSNDPRC54}): ``The efficiency for producing and detecting a 2.2~MeV correlated \gam\ within 2.5 m, with 21--50 PMT hits, and within 1~ms was determined to be $63 \pm 4$\% using the solid curve of Fig. 5. This efficiency is the product of the probability that the \gam\ trigger is not vetoed by a veto shield signal within the previous 15.2~$\mu$s ($82 \pm 1$\%), the data acquisition livetime ($94 \pm 3$\%, lower for $\gamma$'s than for primary events), the requirement that the \gam\ occurs between 8~$\mu$s and 1000~$\mu$s after the primary event ($95 \pm 1$\%), the requirement that the \gam\ has between 21 and 50 hit PMTÕs ($90 \pm 4$\%), and the requirement that the \gam\ reconstructs within 2.5 m of the primary event ($96 \pm 2$\%)''. From this we can conclude that the efficiency for correlated $\gamma$'s stated in LSND's final physics paper, 0.51 ($\pm 7$\%) for $R_{\gamma} > 1$, is in part determined by data acquisition properties, and in part by the `base distributions' of $\Delta r$, $\Delta t$ and
$N_{\rm hits}$. 

While we must assume that the data acquisition properties such as livetime, 
and experimental conditions such as the rate of accidental $\gamma$'s, were understood and correctly taken into account,
we shall dispute below some of the `base distributions' and their impact on the `efficiencies'.

\section{On uncertainties of the `base distributions'}
\label{uncertainties}

In this section, we shall assess systematic uncertainties of the `base distributions' that were published by 
LSND~\cite{LSNDPRD64}  and reproduced in Fig.~\ref{Fig10inPRD64}.

Why did LSND publish smooth functions in their final physics paper~\cite{LSNDPRD64} and not data and Monte Carlo distributions, respectively, as they did in their earlier paper on intermediate results~\cite{LSNDPRC54}? This question is the more appropriate as there are interesting differences between the two presentations, not discussed by LSND anywhere in their papers, to which we shall return below. 

We note also that systematic uncertainties of the `base distributions' are nowhere discussed in LSND papers. 

Our discussion will concentrate on LSND's final base distributions since these were claimed to underlie their final results. We note that the distributions shown in Fig.~\ref{Fig10inPRD64} are computer-drawn functions. We have reproduced the LSND `base distributions' with a plot digitization program. They are referred to below as 
HARP--CDP `base distributions' and used to emulate the \Rgam\ fit hypotheses for correlated and 
accidental $\gamma$'s.

\subsection{The $\Delta t$ `base distributions'}
\label{Deltatdistribution}

$\Delta t$ is the time delay between the primary electron and the subsequent correlated or accidental $\gamma$. 
The $\Delta t$ `base distributions' for 8~$\mu$s $\leq$ $\Delta t$ $\leq$ 1~ms are uncontroversial, and so are the pertinent contributions to the `efficiencies' of correlated and accidental $\gamma$'s.

The $\Delta t$ distribution of accidental $\gamma$'s is flat since we understand that any number of $\gamma$'s was recorded that fell into the window 8 $\mu$s $\leq \Delta t \leq$ 1~ms (the choice of the one with the largest \Rgam\ among them does not depend on $\Delta t$). Because of this requirement---that is nowhere discussed in the LSND papers---, it would have been worthwhile to demonstrate experimentally that the $\Delta t$ distribution of accidental $\gamma's$ was indeed flat. 

\begin{figure*}
\begin{center}
\includegraphics[width=0.8\textwidth]{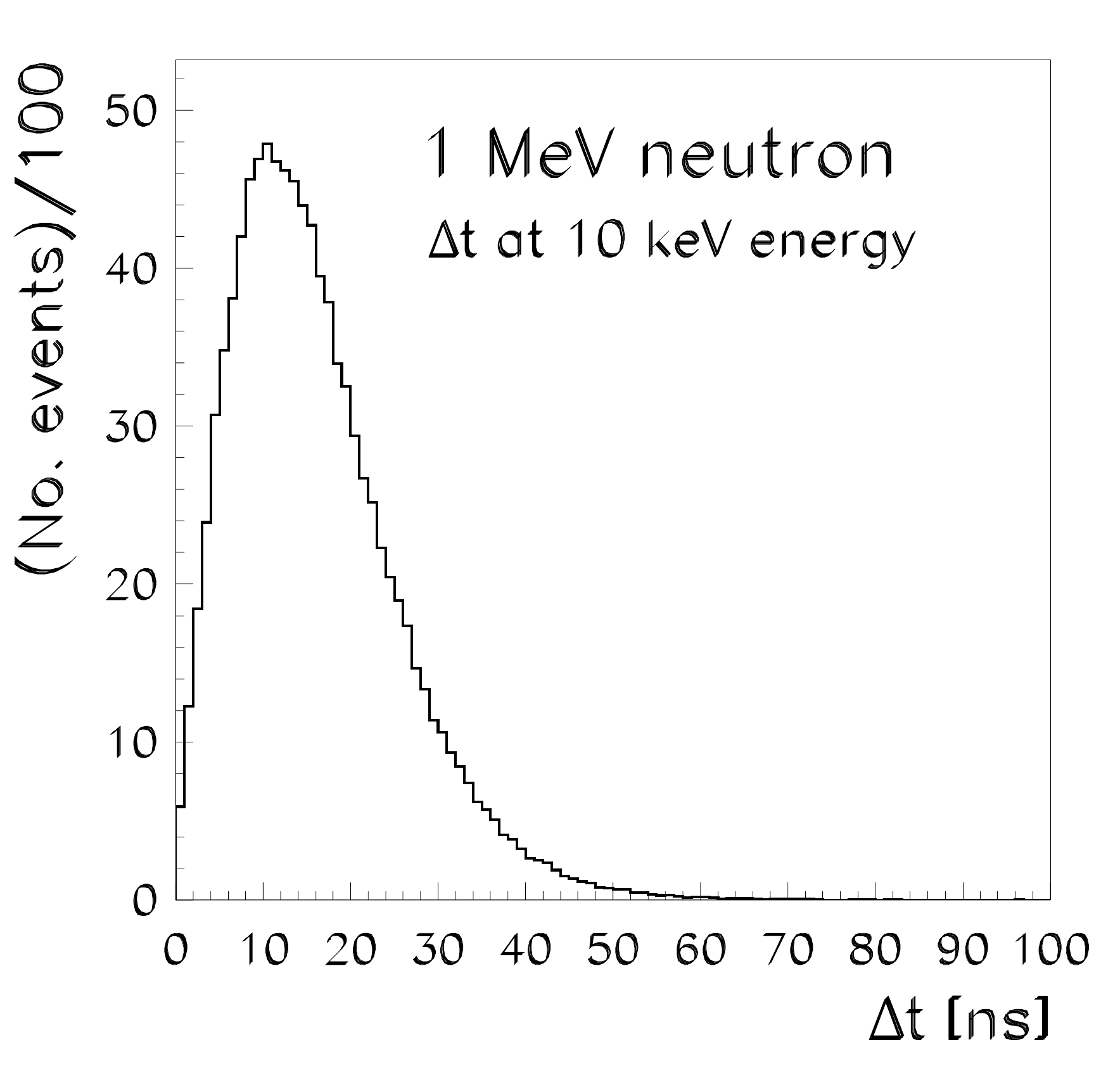}    
\caption{Distribution of time delays between the creation in the LSND scintillator medium of a 1~MeV neutron and the transition to the epithermal energy of 10~keV.}
\label{Deltatatepithermal}
\end{center}
\end{figure*}

\begin{figure*}
\begin{center}
\includegraphics[width=0.8\textwidth]{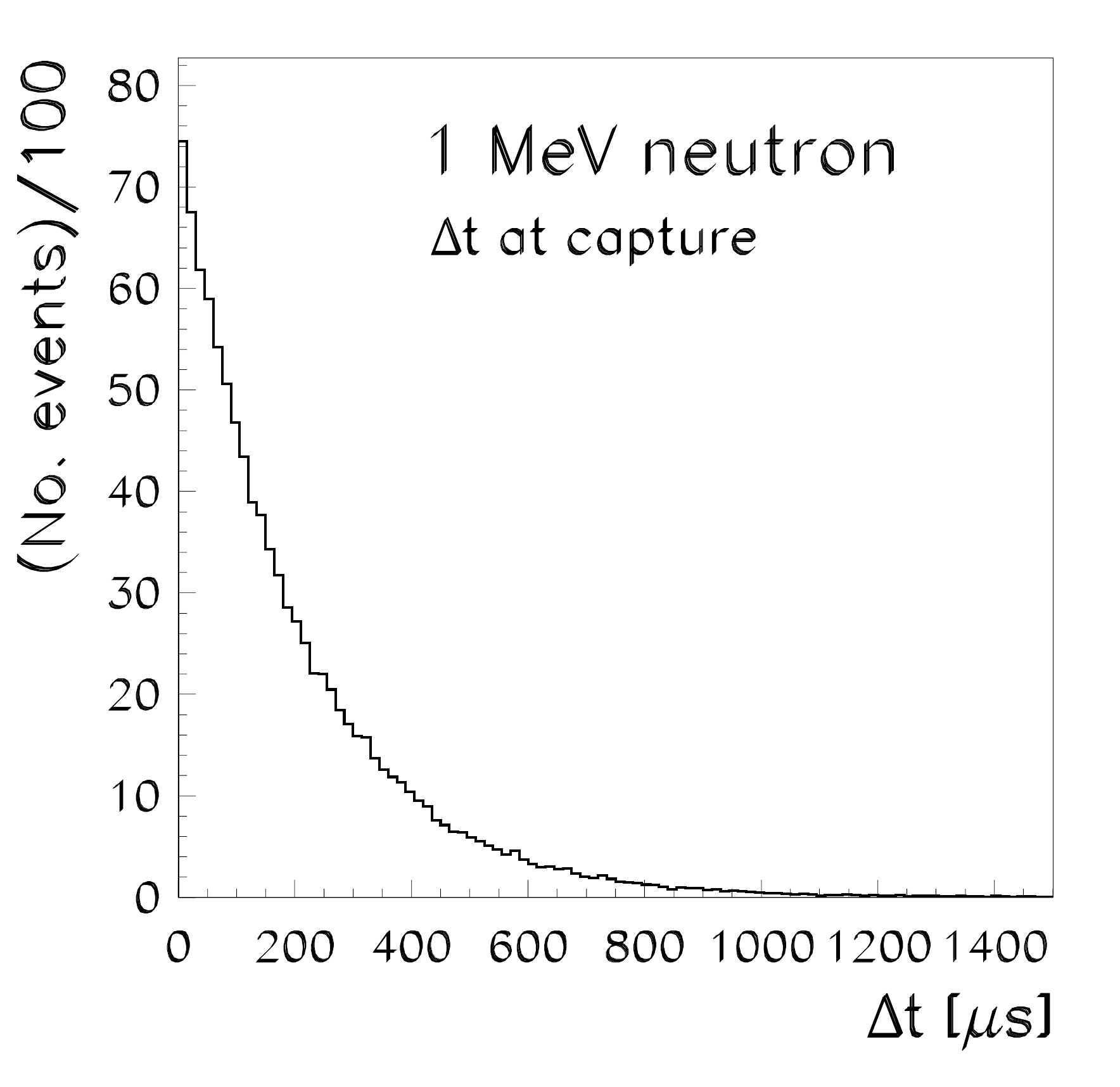}    
\caption{Distribution of time delays between the creation in the LSND scintillator medium of a 1~MeV neutron and its capture on a free proton.}
\label{Deltatatcapture}
\end{center}
\end{figure*}

The $\Delta t$ distribution of
correlated $\gamma$'s is claimed by LSND as exponentially falling with a mean time delay of 186~$\mu$s. 
We checked the latter with a Monte Carlo program that tracks 1~MeV neutrons (the average energy of neutrons released in the signal reaction \signalreaction ) from creation until capture by a free proton in the 
LSND detector\footnote{We are indebted to K.~N\"{u}nighoff from the Forschungszentrum J\"{u}lich GmbH for making available to us an extensive compilation of neutron cross-sections on free protons.}.
Salient results are shown in Figs.~\ref{Deltatatepithermal} and \ref{Deltatatcapture}.
The average time to reach the epithermal energy of 10~keV is 16.0~ns and hence negligibly small in comparison to the average time of 15.0~$\mu$s to reach the thermal energy of 0.022~eV. The bulk of the time is spent between subsequent elastic scatterings at thermal energy. We find the
time delay distribution until capture exponentially falling with an average of 196.1~$\mu$s, slightly larger
than 186~$\mu$s.

This time delay solely depends on the average speed of a thermal neutron and its capture cross-section.
At thermal energy level, the hydrogen atoms in the LSND detector medium, mineral oil, are not free but
bound which increases the capture cross-section with respect to the one for free hydrogen atoms. This 
increase---which entails a decrease of the average time delay---depends on the energy levels of vibrational states of the hydrogen atom in the mineral oil molecule and is hard to estimate. 
LSND do not state where their expectation of 186~$\mu$s comes from, but they state in an earlier paper  that they measured
the average time delay as  $188 \pm 3$~$\mu$s (Section III.B and Fig.~1 in Ref.~\cite{LSNDPRC54}) with
cosmic-ray neutrons. This is possible since the neutron's 
$\Delta t$ distribution is nearly independent of the initial neutron 
energy (other than the neutron's 
$\Delta r$ distribution, discussed below). We acknowledge
this measurement, conclude that the hydrogen atom in mineral oil is quasi-free, and use LSND's average time delay of 186~$\mu$s in our analysis.

\subsection{The $\Delta r$ `base distributions'}

Other than LSND's  $\Delta t$ `base distributions', their $\Delta r$ `base distributions' cause major concerns.

In an infinite medium, the $\Delta r$ distribution of accidental $\gamma$'s should 
smoothly rise with $(\Delta r)^2$.
In the finite fiducial volume of the LSND detector, the actual distribution must fall below this functional dependence,
especially at large $\Delta r$. Indeed, this is seen in Fig.~2 in Ref.~\cite{LSNDPRC54} which shows measured data, but is not seen in Fig.~10 in LSND's final physics paper~\cite{LSNDPRD64} (reproduced 
here in Fig.~\ref{Fig10inPRD64}) which shows not the experimental data but a parametrization obtained from a fit to experimental data. We consider that at large $\Delta r$ LSND's parametrization of the $\Delta r$ distribution of accidental $\gamma$'s is unphysical.

There is a second reason why LSND's final $\Delta r$ distribution of accidental $\gamma$'s cannot be quite right: the `hot spot' of radioactivity in the upstream bottom portion of the LSND detector, prominently visible in Fig.~3 in  
Ref.~\cite{LSNDPRC54} and discussed there in Section III.D.2. Yet there is no mention of this `hot spot' in LSND's final physics paper~\cite{LSNDPRD64}.

\begin{figure*}
\begin{center}
\includegraphics[width=0.8\textwidth]{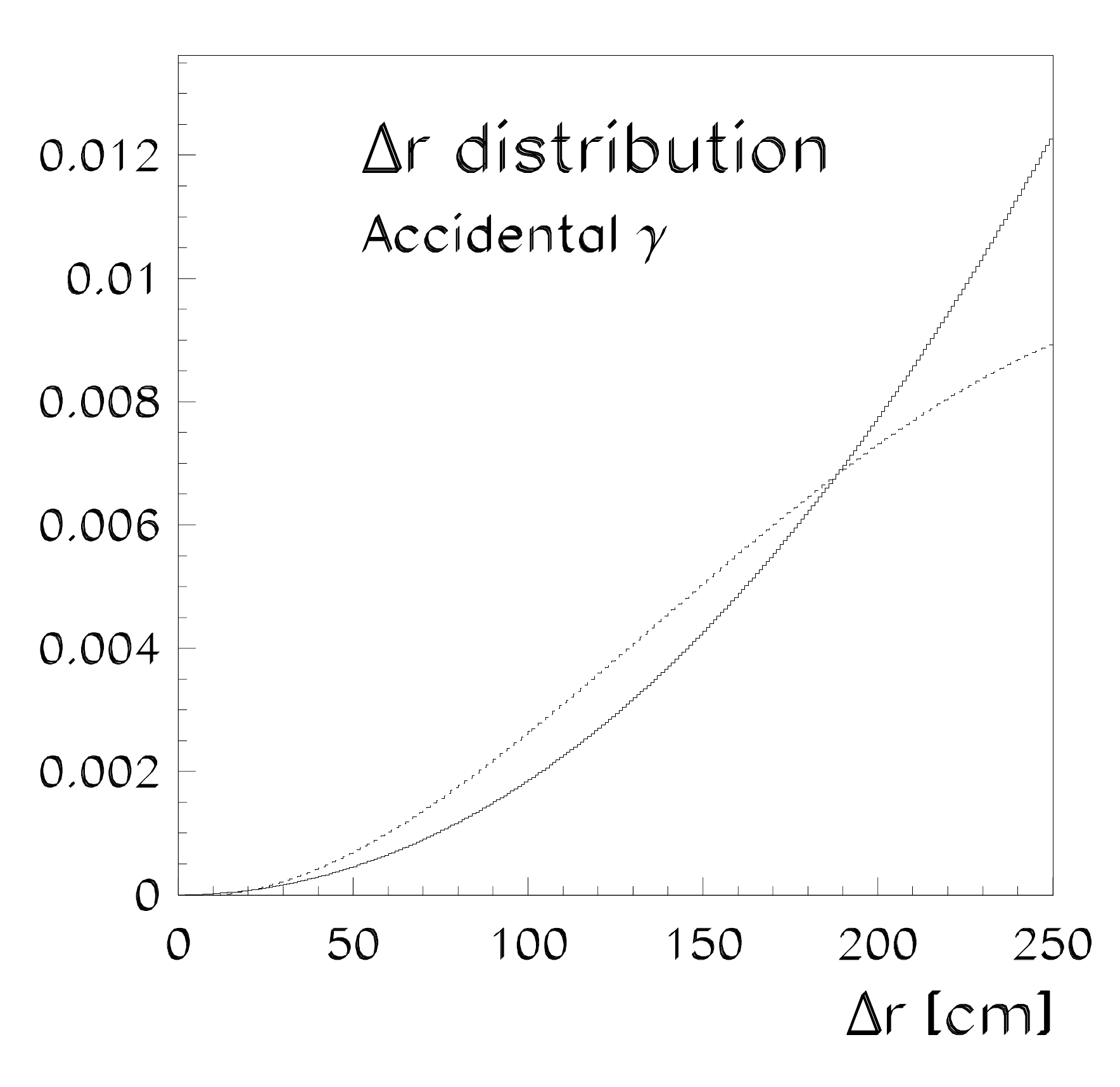}    
\caption{Normalized distributions of the distance $\Delta r$ for accidental $\gamma$'s: as published by LSND (full line), and a variant (dotted line).}
\label{Deltaraccidentalgamma}
\end{center}
\end{figure*}

LSND's $\Delta r$ distribution for accidental $\gamma$'s is compared in Fig.~\ref{Deltaraccidentalgamma} with a variant that we consider equally likely to represent the situation. 

The $\Delta r$ distribution for accidental $\gamma$'s is rather insensitive to the spatial resolution of the
reconstructed position of the accidental $\gamma$. The variant shown in Fig.~\ref{Deltaraccidentalgamma} has been generated with a spatial resolution of $\sigma = 35$~cm.

The $\Delta r$ distribution of accidental $\gamma$'s could experimentally be well determined by recording 
$\gamma$-compatible events in randomly opened gates, and calculating $\Delta r$ w.r.t. a randomly chosen location within the fiducial volume. It would have been worthwhile to present the result in LSND's final physics paper.

The cut $\Delta r \leq 250$~cm that contributes to the `efficiency' of accidental $\gamma$'s will have a different effect for the two distributions shown in Fig.~\ref{Deltaraccidentalgamma}.

Now we turn to the $\Delta r$ distribution of correlated $\gamma$'s. This distribution stems from a convolution of (i) the distribution of the distance of the neutron emitted in the reaction \signalreaction\ between its point of creation and its point of  capture, (ii) the spatial resolution of the reconstructed point of
creation, and (iii) the spatial resolution of the reconstructed position of the 2.2~MeV \gam . 

\begin{figure*}
\begin{center}
\includegraphics[width=0.8\textwidth]{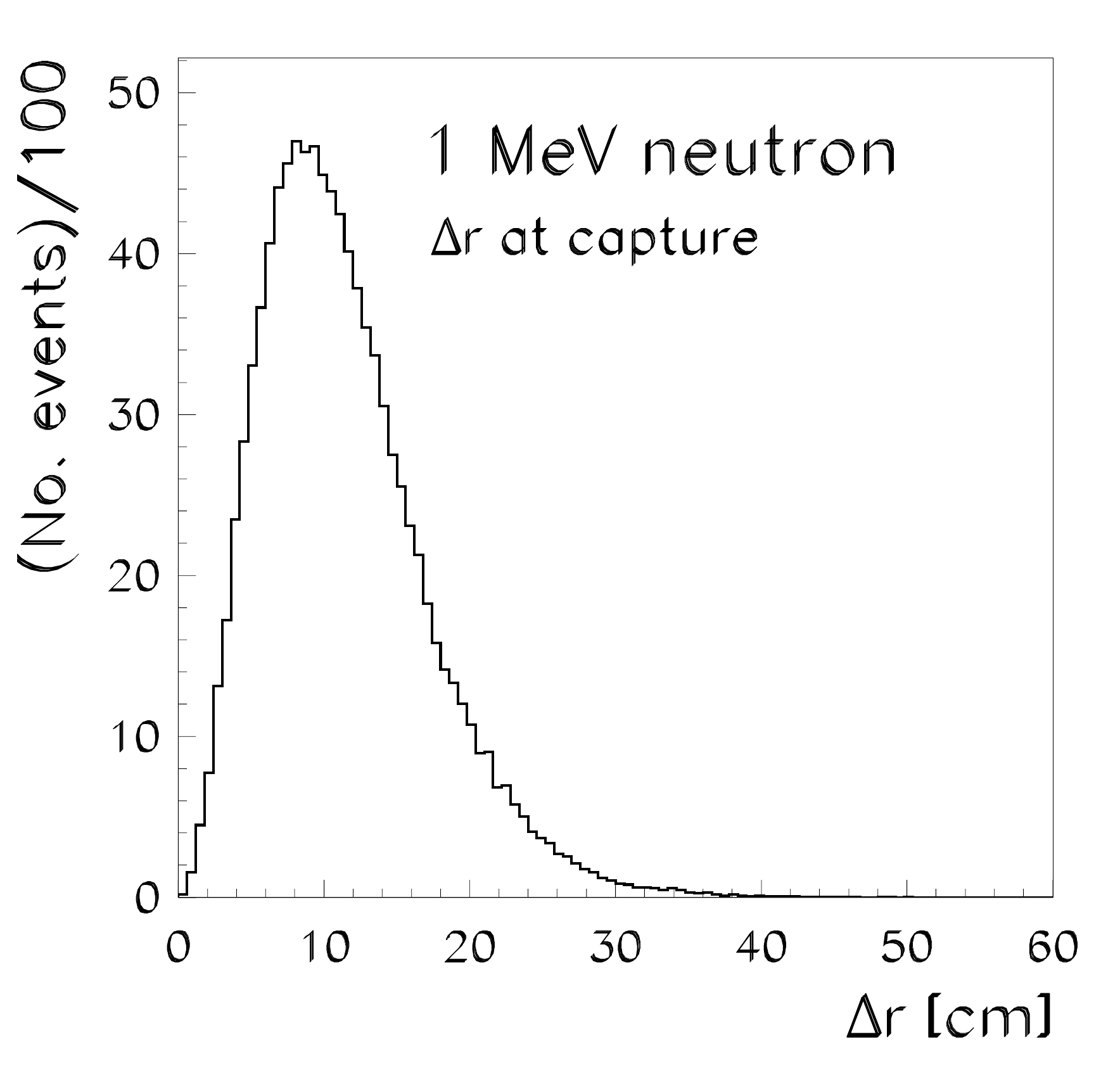}    
\caption{Distribution of distances between the point of creation in the LSND scintillator medium of a 1~MeV neutron and the point of capture on a free proton.}
\label{Deltaratcapture}
\end{center}
\end{figure*}

Our result on the first of these three distributions is shown in Fig.~\ref{Deltaratcapture}. The average $\Delta r$ between the point of creation of a 1~MeV neutron and the point of its
capture is 11.4~cm. This is consistent with expectation from neutron diffusion theory which stipulates that thermal neutrons are captured at an average distance 
of $2 \sqrt{D_{\rm diff} \cdot \lambda_{\rm abs}}  \sim 4.7$~cm, were
$D_{\rm diff} = 0.144$~cm is the neutron diffusion coefficient and $\lambda_{\rm abs} = 38.4$~cm
is the mean free path in mineral oil. While these numbers hold in the limit of free protons, we recall our conclusion from Section~\ref{Deltatdistribution} that the hydrogen atom in mineral oil is quasi-free. The increase from
4.7~cm to 11.4~cm stems from the neutron's movement during slowing down from 1~MeV kinetic energy
to the average thermal energy of 0.022~eV. We note that there is no discussion in the LSND papers 
of the distribution of the distance $\Delta r$  of the neutron between creation and capture. 

The point of neutron creation is reconstructed as the position of the primary electron. LSND state the
average spatial resolution as 14~cm (Section II.E in Ref.~\cite{LSNDPRD64}).

The spatial resolution of 2.2~MeV $\gamma$'s is recognized to be important by LSND, however they do not
specify the resolution. The only statement that can be found reads as ``the most likely distance was reduced from 74~cm [with a previous reconstruction algorithm] to 55~cm'' (Section IV in Ref.~\cite{LSNDPRD64}). A peak position of 55~cm suggests a spatial resolution of approximately 35~cm when taking the said convolution into account. This is considerably smaller than the estimate of 54~cm which is 14~cm multiplied by $\sqrt{33/2.2}$, where 33~MeV is the average primary electron energy above the threshold of 20~MeV. So there is
quite some uncertainty on the spatial resolution of 2.2~MeV $\gamma$'s. 

We note that all three distributions that are convoluted into the $\Delta r$ distribution of
correlated $\gamma$'s, come from Monte Carlo simulation and cannot be verified by data in an unbiased
manner. In particular, cosmic-ray neutrons have much higher energy and bias 
$\Delta r$ towards larger values by virtue of the decreasing scattering cross-section above 1~MeV energy. 
For the 2.2~MeV photons from the capture of cosmic-ray neutrons, there is no unbiased reference 
point for the calculation of $\Delta r$.

\begin{figure*}
\begin{center}
\includegraphics[width=0.8\textwidth]{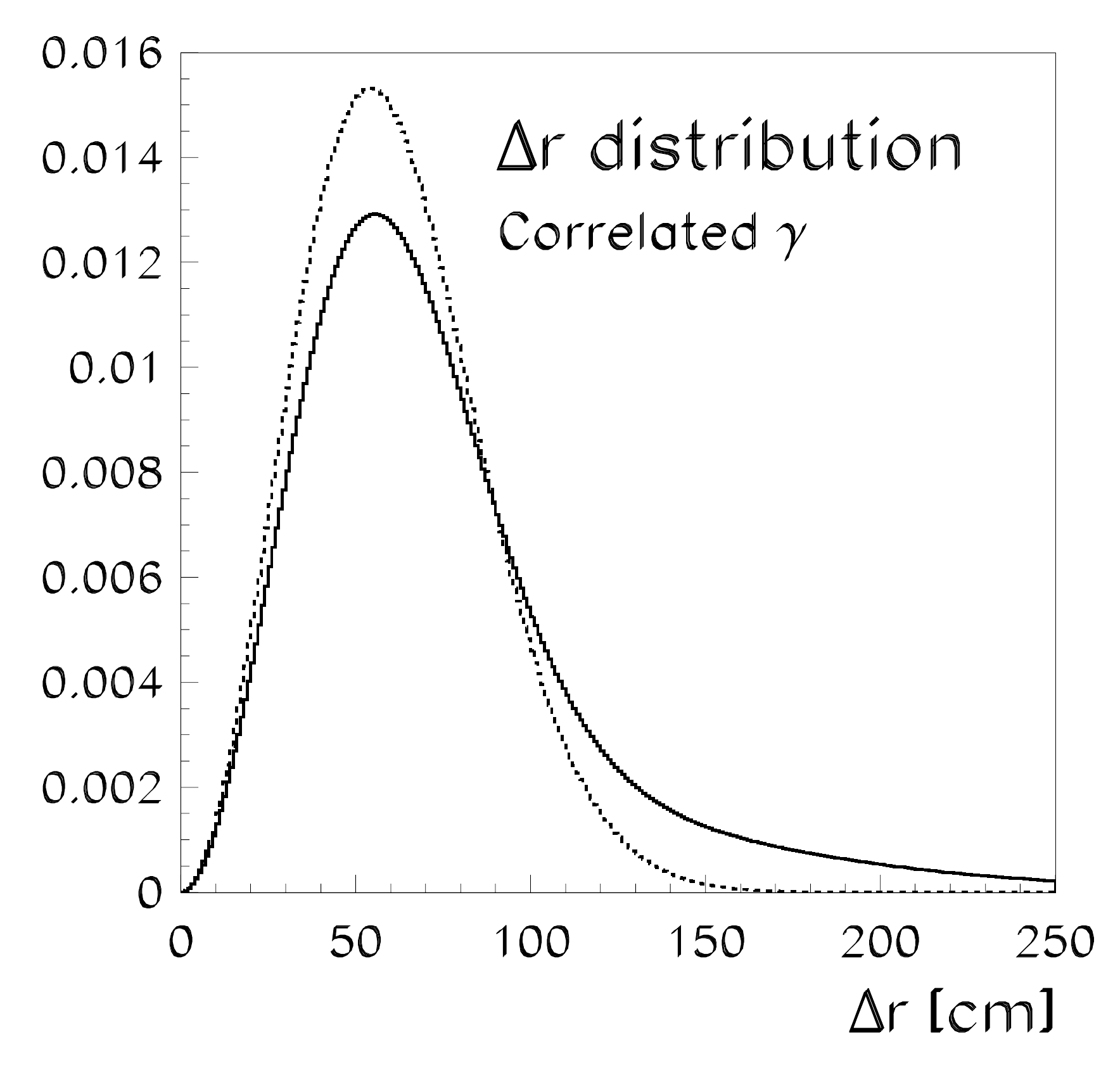}    
\caption{Normalized distributions of the distance $\Delta r$ for correlated $\gamma$'s: as published by LSND (full line), and a variant (dotted line).}
\label{Deltarcorrelatedgamma}
\end{center}
\end{figure*}

We compare in Fig.~\ref{Deltarcorrelatedgamma} the 
$\Delta r$ distribution claimed by LSND with with a variant that we consider 
equally likely to represent the situation. 

As both the LSND $\Delta r$ distribution for correlated $\gamma$'s and our respective distribution
tend toward zero at large $\Delta r$, the cut  $\Delta r \leq 250$~cm will not cause much difference on the `efficiency' of correlated $\gamma$'s.

\subsection{The $N_{\rm hits}$ `base distributions'}

While the $\Delta r$ distributions of correlated and accidental $\gamma$'s cause major concerns, the $N_{\rm hits}$ distributions cause minor concerns.
This is because the pertinent $N_{\rm hits}$ distribution can be experimentally verified with high statistical 
precision from the measurement of 2.2~MeV $\gamma$'s from the capture of cosmic-ray neutrons\footnote{Clean samples of 2.2~MeV $\gamma$'s from neutron capture could be ascertained through the observation of an exponentially falling $\Delta t$ distribution with an average of 186~$\mu$s.}, and from the measurement of accidental 
$\gamma$'s in randomly opened gates during beam-off times.

LSND state that accidental $\gamma$'s arise from ``radioactivity''. We note that 
that there are small contributions from misidentified electrons from $^{12}$B and positrons 
from $^{12}$N$_{\rm gs}$ beta decays which both exhibit a $N_{\rm hits}$ spectrum that is quite different from
the $N_{\rm hits}$ spectrum from radioactivity. This is well visible in the respective measured 
spectrum shown in Fig.~2 in Ref.~\cite{LSNDPRD64}, however, the tail toward large $N_{\rm hits}$ that is expected and well visible there, is absent in Fig.~\ref{Fig10inPRD64} taken from the final LSND physics paper~\cite{LSNDPRD64}. Another concern
is the shape of the spectrum toward small $N_{\rm hits}$ which is dominated by a strong non-linearity of the
readout electronics\footnote{We note the stark conflict between the calibration 21 PMT hits = 0.7~MeV stated in Section II.F in Ref.~\cite{LSNDPRD64}, and 34~PMT hits = 2.2~MeV read off from Fig.~\ref{Fig10inPRD64}.},
and hence quite vulnerable to even small drifts of the electronics which are unavoidable 
over six years of data taking.

\begin{figure*}
\begin{center}
\includegraphics[width=0.8\textwidth]{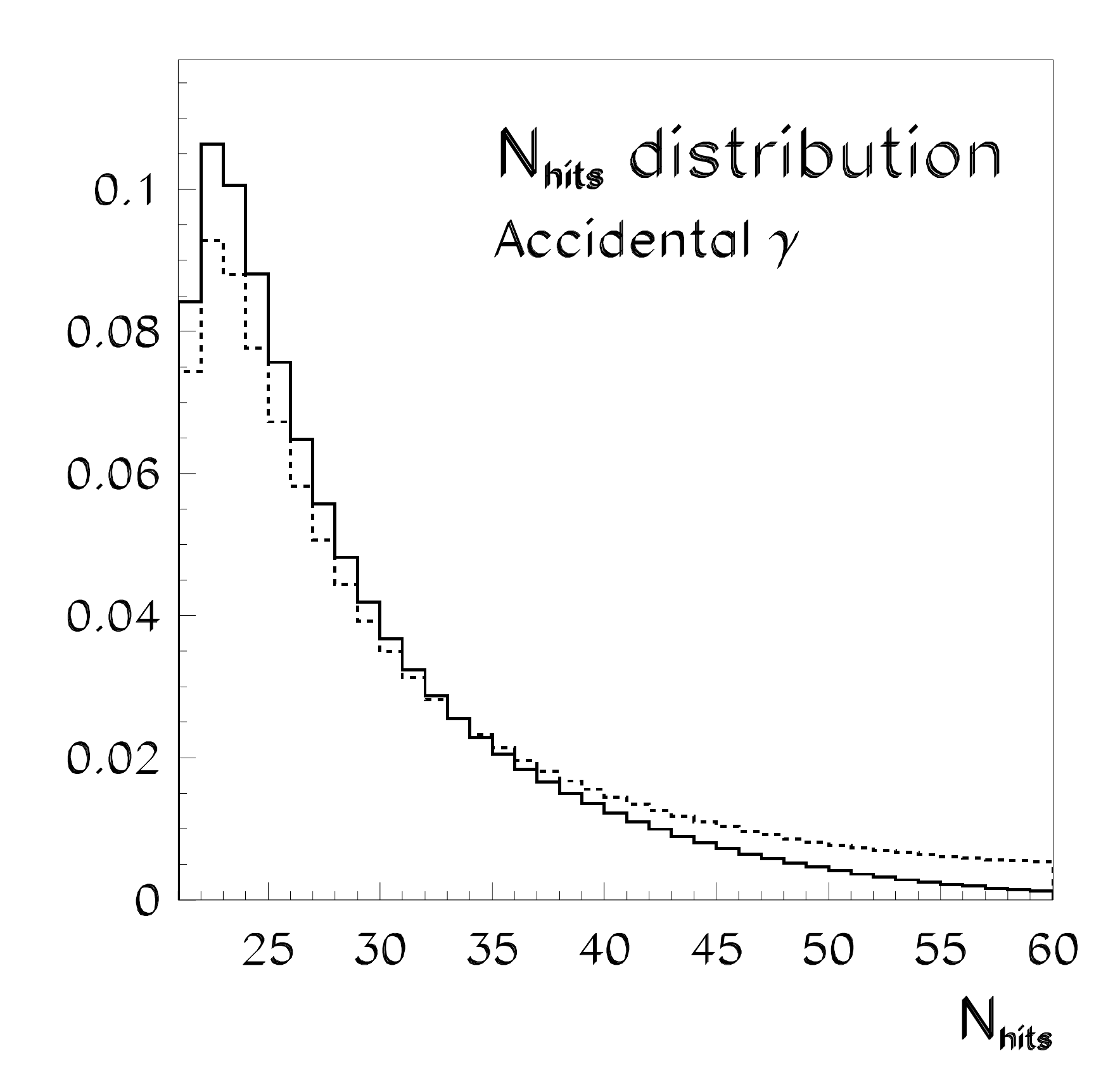}    
\caption{Normalized distributions of the number of hit PMTs, $N_{\rm hits}$, for accidental $\gamma$'s: as published by LSND (full line), and a variant (dotted line).}
\label{Nhitsaccidentalgamma}
\end{center}
\end{figure*}

\begin{figure*}
\begin{center}
\includegraphics[width=0.8\textwidth]{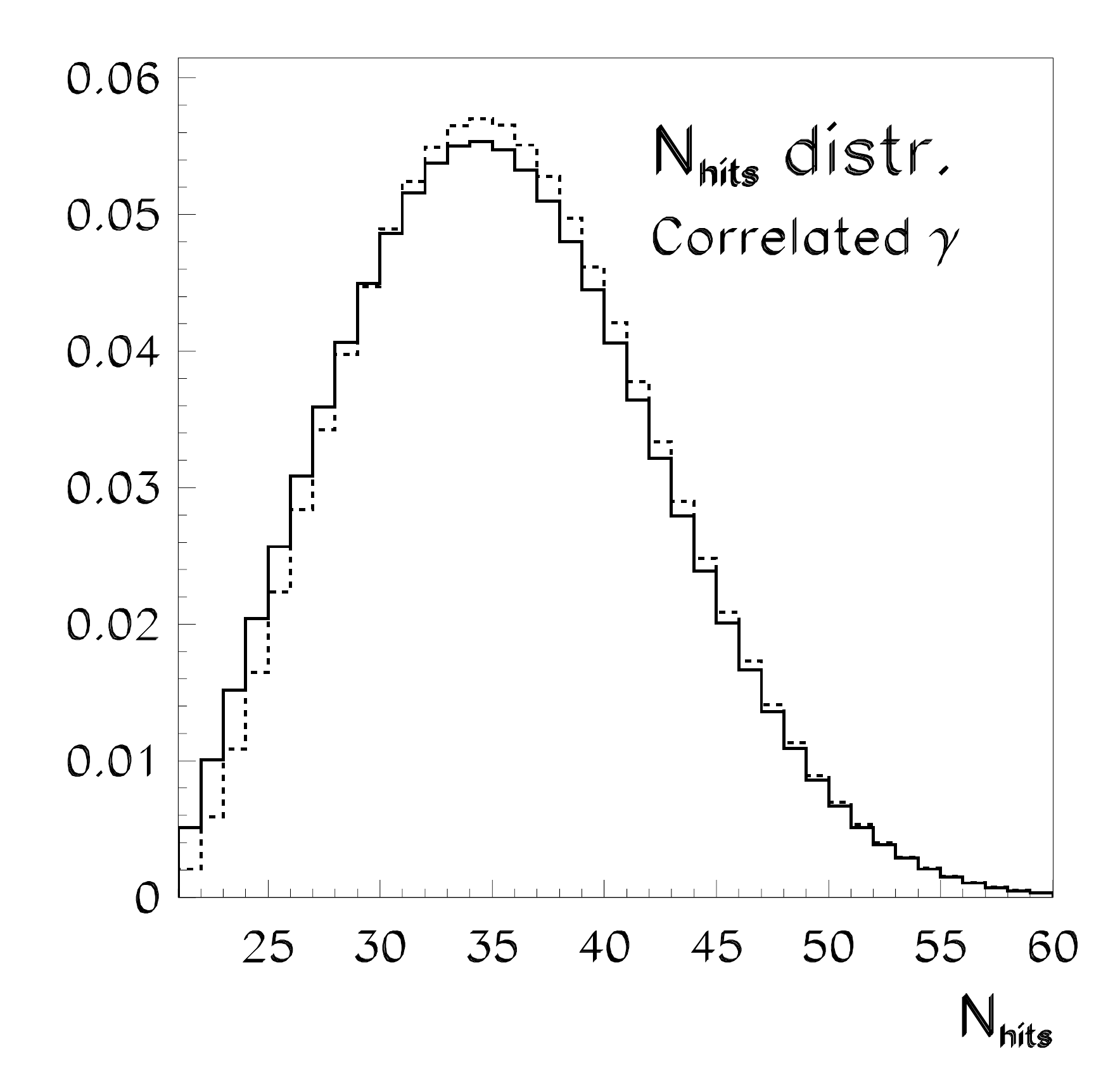}    
\caption{Normalized distributions of the number of hit PMTs, $N_{\rm hits}$, for correlated $\gamma$'s: as published by LSND (full line), and a variant (dotted line).}
\label{Nhitscorrelatedgamma}
\end{center}
\end{figure*}

LSND's $N_{\rm hits}$ distribution for accidental $\gamma$'s is compared in Fig.~\ref{Nhitsaccidentalgamma} with a variant that we consider equally likely to represent the situation. Analogously, Fig.~\ref{Nhitscorrelatedgamma} shows the situation for correlated $\gamma$'s.

We note that the lower cutoff of 21~$N_{\rm hits}$ bears on the `efficiencies' of correlated and even more of accidental $\gamma$'s.

\section{On the `efficiencies' of correlated and accidental $\gamma$'s}
\label{efficienciesofgammas}

As pointed out already above, the numerical values and the uncertainties of the `efficiencies' of correlated and accidental $\gamma$'s are important ingredients to the fit of the \Rgam\ distribution that yields the `beam excess'. After we assessed the `base distributions' and discussed their impact on the `efficiencies', we can give our numerical estimates, which are summarized and compared with the respective values quoted by LSND~\cite{LSNDPRD64} 
in Table~\ref{efficiencies}.
\begin{table*}[htdp]
\caption{Our estimate of the `efficiencies' of correlated and accidental $\gamma$'s}
\begin{center}
\begin{tabular}{|l|c|c|}
\hline
     &  Correlated $\gamma$'s &  Accidental $\gamma$'s  \\
\hline
\hline
No veto signal   & \multicolumn{2}{|c|}{0.82 $(\pm 1\%)$} \\                                    
Data acquisition alive    & \multicolumn{2}{|c|}{0.94 $(\pm 3\%)$} \\
\hline
$\Delta r \leq 250$~cm   &  0.98 $(\pm 2\%)$  &  0.23 $(\pm 4\%$)           \\ 
8~$\mu$s $ \leq \Delta t \leq $ 1~ms  & 0.95 $(\pm 1\%)$  &  0.63 $(\pm ?\%)$   \\
$21 \leq N_{\rm hits} \leq 60$ &  0.94 $(\pm 2\%)$  &  0.85 $(\pm 5\%)$    \\
\hline
\hline
Overall `efficiency'  for $R_{\gamma} > 0$ &        0.67 $(\pm 4\%)$      &     0.095   $(\geq \pm 7 \%)$        \\
($R_{\gamma} > 1$) / ($R_{\gamma} > 0$)     &    0.93          &    0.072           \\
Overall `efficiency' for $R_{\gamma} > 1$  &        0.63  $(\pm 4\%)$     &     0.0068  $(\geq \pm 7 \%)$       \\
\hline
\hline
LSND's `efficiency' for $R_{\gamma} > 1$  &   0.51 $(\pm 7\%)$  & 0.012 $(\pm 7\%)$ \\  
\hline
\end{tabular}
\end{center}
\label{efficiencies}
\end{table*}

Several comments are in order. First, the stated `efficiencies' from the signal veto and data acquisition livetime
are not stated in LSND's final physics paper~\cite{LSNDPRD64} but are taken from an earlier paper that gave
intermediate results (Section III.F in Ref.~\cite{LSNDPRC54}), and hence may not be correct averages for the
full data set. Second, our estimated `efficiency' of correlated $\gamma$'s for $R_{\gamma} > 1$ of 0.63 ($\pm 4\%$) does not compare well with LSND's which is 0.51 ($\pm 7\%$). This is important as the `beam excess' is directly
proportional to the reciprocal value of this `efficiency'. Third, an important information is missing in LSND's final physics paper~\cite{LSNDPRD64} which is the average rate of accidental $\gamma$'s over the entire data taking which directly bears on the efficiency
of observing at least one accidental \gam\ in the window 8~$\mu$s $ \leq \Delta t \leq $ 1~ms. The only pertinent information is found in an earlier paper that gave intermediate results where it is stated ``The average accidental \gam\ rate over the entire detector is $1.07 \pm 0.01$~kHz in 1993, $1.19 \pm 0.01$~kHz in 1994, and $1.14 \pm 0.01$~kHz in 1995'' (Section III.D.2 in Ref.~\cite{LSNDPRC54}). We used an average accidental rate after deadtime of 1~kHz
which leads to the respective `efficiency' of 0.63 quoted in Table~\ref{efficiencies} for the window 8~$\mu$s $ \leq \Delta t \leq $ 1~ms. We have no explanation why our estimate of the overall `efficiency' for $R_{\gamma} > 1$
of 0.0068 $(\geq \pm 7 \%)$ is nearly a factor of two below LSND's estimate of 0.012 $(\pm 7\%)$. We recall in this context that the `efficiencies' and their uncertainties are quoted but not explained in any way in LSND's final physics paper~\cite{LSNDPRD64}, despite their importance.

While we are concerned about these inconsistencies, for lack of published information on details of the LSND experiment we cannot insist on our estimates of `efficiencies'. Hence we shall use in the further discussion the values
and uncertainties quoted by LSND.

\section{Positrons from $^{12}$N$_{\rm gs}$ beta decays}
\label{positrons}

By virtue of its construction, the LSND detector could only measure the
position and time of an energy deposit, but could not determine 
whether the deposit originates from a charged or from a neutral particle. 

In the same way as electrons from (accidental) beta 
decays of $^{12}$B nuclei---activated by the capture of cosmic-ray \mum\ 
by $^{12}$C nuclei---contribute to uncorrelated $\gamma$'s,
positrons from the abundant beta decay of $^{12}$N$_{\rm gs}$ 
nuclei originating from the inverse beta decay reaction
$\nu_{\rm e}$ + $^{12}$C $\rightarrow$ e$^-$ + $^{12}$N$_{\rm gs}$ can 
contribute to beam-related $\gamma$'s.

Because of the threshold energy of 17.86~MeV for the reaction
$\nu_{\rm e}$ $^{12}$C $\rightarrow$ e$^-$ $^{12}$N 
and the trigger requirement of a minimum electron energy of 20~MeV, only
\nue\ above 37.86~MeV will contribute to background.

The beta decay $^{12}$N$_{\rm gs}$  $\rightarrow$ $^{12}$C e$^+$ $\nu_{\rm e}$ proceeds with
$E_{\rm max} = 16.316$~MeV and a lifetime of 15.87~ms~\cite{Ajzenberg1990}.    

Some positrons 
will fake a \gam\ from neutron capture. The fraction of positrons with
accepted energy between 21 and 60 PMT hits (corresponding to an energy
calibration of 34~PMT hits = 2.2~MeV read off from Fig.~\ref{Fig10inPRD64}),
is calculated to be 0.068 (for the calculation of this fraction, the
addition of 1.022~MeV from positron annihilation and the energy resolution have
been taken into account). The window 8~$\mu$s $ \leq \Delta t \leq $ 1~ms selects
a fraction of 0.061 of the positrons.

Altogether, we estimate that 2.7~positrons meet the acceptance criteria of a \gam\ from neutron capture,
with an estimated `efficiency' (defined analogously to the `efficiency' of $\gamma$'s) of 0.58. 
Their $\Delta r$, $\Delta t$ and $N_{\rm hits}$ distributions resemble more closely the ones of
correlated $\gamma$'s rather than the ones of accidental $\gamma$'s. Therefore, misidentified positrons from
$^{12}$N$_{\rm gs}$ beta decays constitute a small but non-negligible background.

This background of misidentified positrons from
$^{12}$N$_{\rm gs}$ beta decays is not considered in the LSND analysis.

\section{Pseudodata {\em in lieu\/} of LSND data}
\label{pseudodata}

\subsection{The methodology}

The LSND analysis rests on the following concept. Each accepted primary electron candidate is entered into the \Rgam\ plot where \Rgam\ is calculated by the \Rgam\ algorithm 
from the observed triplet [$\Delta r$, $\Delta t$, $N_{\rm hits}$] of the associated \gam\ as the 
ratio of the likelihoods that this triplet stems from a
correlated \gam\ and from an accidental \gam , respectively. The \Rgam\ algorithm makes use of the 
`base distributions'. The distribution of the $R_{\gamma}$'s  represent the data.  
The same \Rgam\ algorithm is also used to calculate the \Rgam\ distributions of correlated and accidental $\gamma$'s, i.e., of $\gamma$'s of the two possible origins. Both the \Rgam\ distributions of
correlated and accidental $\gamma$'s are normalized to unity, their linear combination constitutes the fit hypothesis.
Out of the two fit parameters, the coefficient of the correlated $R_\gamma$ distribution is the wanted `beam excess'.

There is no reason to believe that the `base distributions' are error-free. Rather, they can vary within 
bands of systematic uncertainty. 
LSND did not publish a study of the systematic error of the `beam excess', they quoted its error without commenting on the origin. 
The usual way of studying the systematic error of the `beam excess' that stems from uncertainties of the
`base distributions' would have been the following. Starting from their
observed triplets  [$\Delta r$, $\Delta t$, $N_{\rm hits}$], they could have varied the `base distributions' used in the $R_\gamma$ algorithm. This would have produced variants of both the $R_\gamma$ distribution of the data and the $R_\gamma$ distributions of correlated and accidental $\gamma$'s. They would have noticed both changes of the `beam excess' and changes of its error (it being understood that `error' refers to the correlated error, i.e., the correlation between fit parameters is duly taken into account).

While this should have been done by LSND but was apparently not---there is no pertinent mention in their papers---, we cannot do that since we have no 
access to the observed triplets  [$\Delta r$, $\Delta t$, $N_{\rm hits}$] .
Therefore, we turn things around. 
First, we generate pseudodata  with variants of `base distributions'. Then, we calculate with an \Rgam\ algorithm that makes use of one and only one set of `base distributions', namely the HARP--CDP `base distributions', both the \Rgam\ distribution of the pseudodata and the fit hypothesis, i.e., the linear combination of the normalized
\Rgam\ distributions of correlated and accidental $\gamma$'s. We recall that the HARP--CDP `base distributions' reproduce the `base distributions' 
published by LSND and shown in Fig.~\ref{Fig10inPRD64}. For each variant of the `base distributions', we
observe a change of the `beam excess' and a change of its error when we compare with the default situation: the pseudodata are generated with the same set of HARP--CDP `base distributions' that the $R_{\gamma}$ algorithm uses to calculate the \Rgam\ distribution of the pseudodata and of the normalized \Rgam\ distributions of correlated and accidental $\gamma$'s.

The observed variation of the `beam excess' that stems from the use of variants of the `base distributions' for the generation of pseudodata is used to obtain an estimate of its systematic error.

We draw attention to a peculiar correlation that is inherent in the LSND analysis procedure. The
very same \Rgam\ algorithm is used to calculate the \Rgam\ distribution of the data and of the normalized \Rgam\ distributions of correlated and accidental $\gamma$'s. That has the following consequence. If the \Rgam\
algorithm makes use of different `base distributions', both the data and the fit hypothesis change coherently.
This manifests itself in a change of the error of the `beam excess' by virtue of a change of the
correlation coefficient of the two fit parameters (the \Rgam\ distributions of correlated and uncorrrelated
$\gamma$'s become more similar or dissimilar). With a view to highlighting this aspect, we list in 
later pseudodata fits in addition to the best-fit value of the `beam excess' and its correlated error also the correlation coefficient. We make no further use of this aspect in our estimate of the 
systematic error of the `beam excess', though.

A problem in our procedure is caused by the presence of accidental $\gamma$'s that occur with a rate that is rather uncertain. Since in the case of more than one accepted \gam\ the one with the largest \Rgam\ is taken, a correct generation of pseudodata depends on the effective rate of accidental $\gamma$'s after data acquisition losses---an information that is not given by LSND. Along the lines of the pertinent discussion in Section~\ref{efficienciesofgammas}, we assume an effective rate (i.e., after data acquisition losses) of accidental $\gamma$'s of 1~kHz and assign a $\pm20\%$ error. 

The event numbers and errors of the pseudodata orient themselves on the respective LSND event numbers 
and errors as listed in Table~\ref{DigitizationofFig14inPRD64}. The errors of the pseudodata are forced to be
\begin{displaymath}
\sqrt{\rm No. \: pseudodata \: events} \cdot \frac{\rm LSND \: data \: error}{\sqrt{\rm No. \: LSND \: events}}
\end{displaymath}

The pseudodata represent expectation values, no fluctuations
from errors are imposed. Thereby, systematic shifts of the `beam excess' in different sets of 
pseudodata are not hidden by statistical variations. 
This permits to demonstrate even small systematic changes of the `beam excess' when input assumptions are varied within uncertainty bounds. 

The discussion of changes of the `beam excess' must occur against a reference point. Closely following LSND's beam excess of $117.9$ events, we chose the round number of 120~events. 

Therefore, our `reference' pseudodata consist of 
\begin{itemize}
\item 120 events with the \Rgam\ distribution of correlated $\gamma$'s derived from the HARP--CDP `base distributions' and with LSND's nominal `efficiency' for $R_{\gamma} > 1$ of 0.51, or suitable variants;
\item 1980 events with the \Rgam\ distribution of accidental $\gamma$'s derived from the HARP--CDP `base distributions' and with LSND's nominal 'efficiency' for $R_{\gamma} > 1$ 
of 0.012, or suitable variants; and
\item (optionally) 2.7~events with the $R_{\beta}$ distribution of positrons from
$^{12}$N$_{\rm gs}$ beta decays (that will be graphically shown below), with an estimated `efficiency' of 0.58.
\end{itemize}

Before presenting results of fits of pseudodata with varying assumptions, we discuss some details of the
HARP--CDP \Rgam\ `reference' distributions that we used. 

\begin{figure*}
\begin{center}
\begin{tabular}{c}
\includegraphics[width=0.6\textwidth]{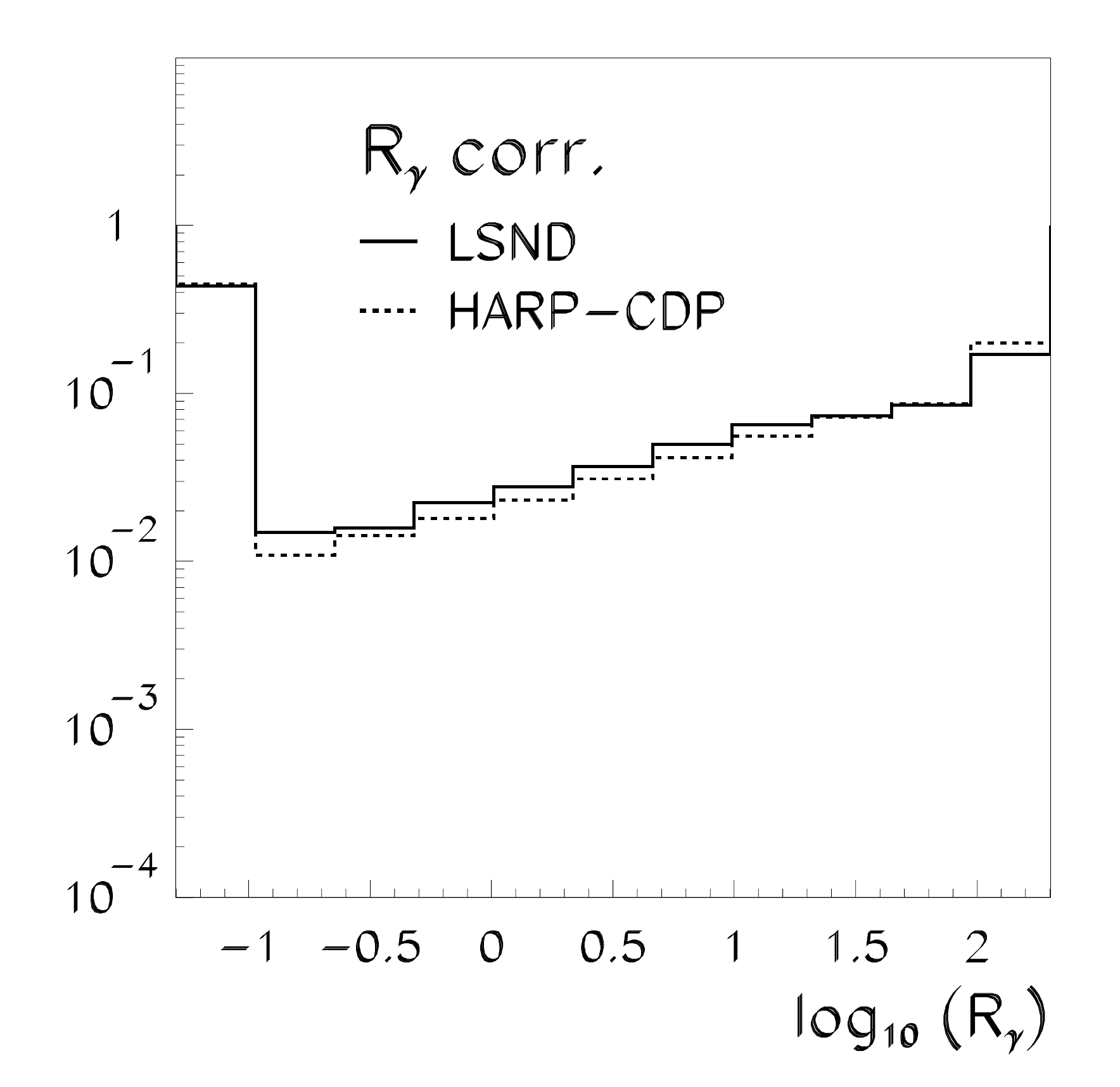} \\   
\includegraphics[width=0.6\textwidth]{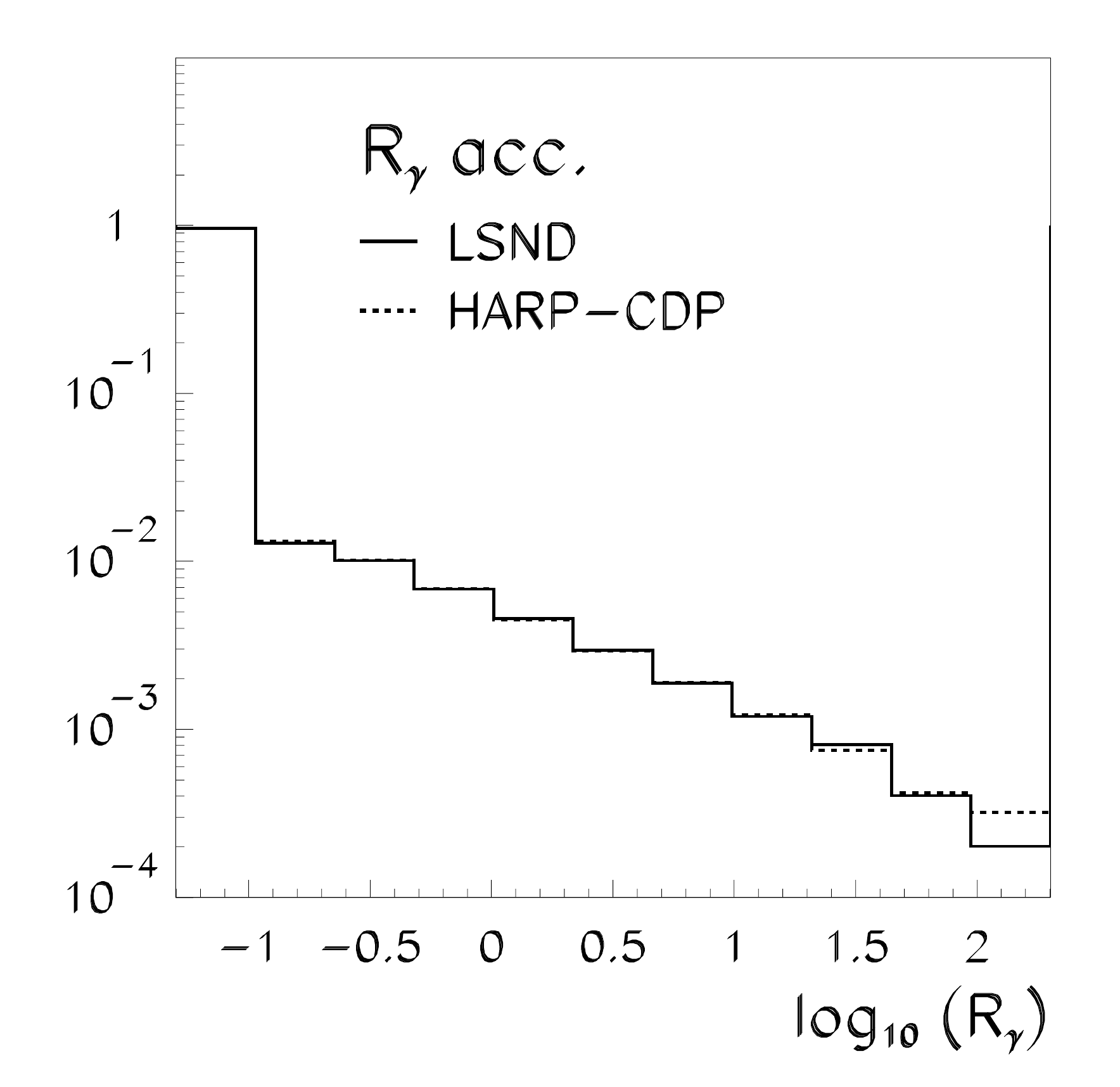} \\ 
\end{tabular}  
\caption{The normalized LSND $R_{\gamma}$ hypotheses (full lines) for correlated (upper panel) and accidental (lower panel)
$\gamma$'s, compared with the respective normalized HARP--CDP $R_{\gamma}$ hypotheses (broken lines); the former are a graphical
representation (duly normalized) of the numbers listed in Table~\ref{DigitizationofFig14inPRD64};  as for way the latter were obtained, see the text.}   
\label{Rgamma-LSNDvsHARPCDP}
\end{center}
\end{figure*}

Figure~\ref{Rgamma-LSNDvsHARPCDP} shows the \Rgam\ hypotheses for correlated and for accidental $\gamma$'s used in the LSND analysis and compares them with the respective HARP--CDP \Rgam\ hypotheses. We recall that the
HARP--CDP \Rgam\ hypotheses stem from the LSND `base distributions' shown in Fig.~\ref{Fig10inPRD64}, modified to take accidental $\gamma$'s with a rate of 1~kHz into account (the size of this modification will be graphically shown below),
and taking the LSND `efficiencies' of 0.51 and 0.012, respectively, for correlated and accidental $\gamma$'s into account. Therefore, the HARP--CDP \Rgam\ hypotheses should be identical to the LSND \Rgam\ hypotheses.
They are not. The effect of the difference for the `beam excess' is shown in 
Table~\ref{FitsofFig14data} where we present the results of both 
an 11-bin (Fit No.~3) and a 10-bin fit (Fit No.~3a), with a view to eliminating concerns from the bin contents 
and error of the first \Rgam\ bin. 

The  `beam excess' of $103.1 \pm 20.2$ events from the 11-bin fit is considerably lower than LSND's `beam excess' of $117.9 \pm 22.4$ (cf. Table~\ref{FitsofFig14data}). The discrepancy is not caused by uncertainties of the
bin contents or a data error of the first \Rgam\ bin since the 10-bin fit gives nearly the same result. 
Rather, the discrepancy proves an inconsistency in the LSND analysis: the \Rgam\ hypotheses used by LSND in the fit of their data and shown in Fig.~\ref{Fig14inPRD64} are not congruent with those derived from LSND's `base distributions' shown in Fig.~\ref{Fig10inPRD64}, although LSND's final physics paper~\cite{LSNDPRD64} claims they are.

We note this discrepancy which has a large effect on the `beam excess'. However, there may well be intrinsic features
of the LSND analysis (such as possible averaging over different `base distributions' at different periods of data taking---albeit not discussed anywhere in the LSND papers---) that we are not aware of. Therefore, we do not make any use of this discrepancy.

\begin{figure*}
\begin{center}
\begin{tabular}{c}
\includegraphics[width=0.6\textwidth]{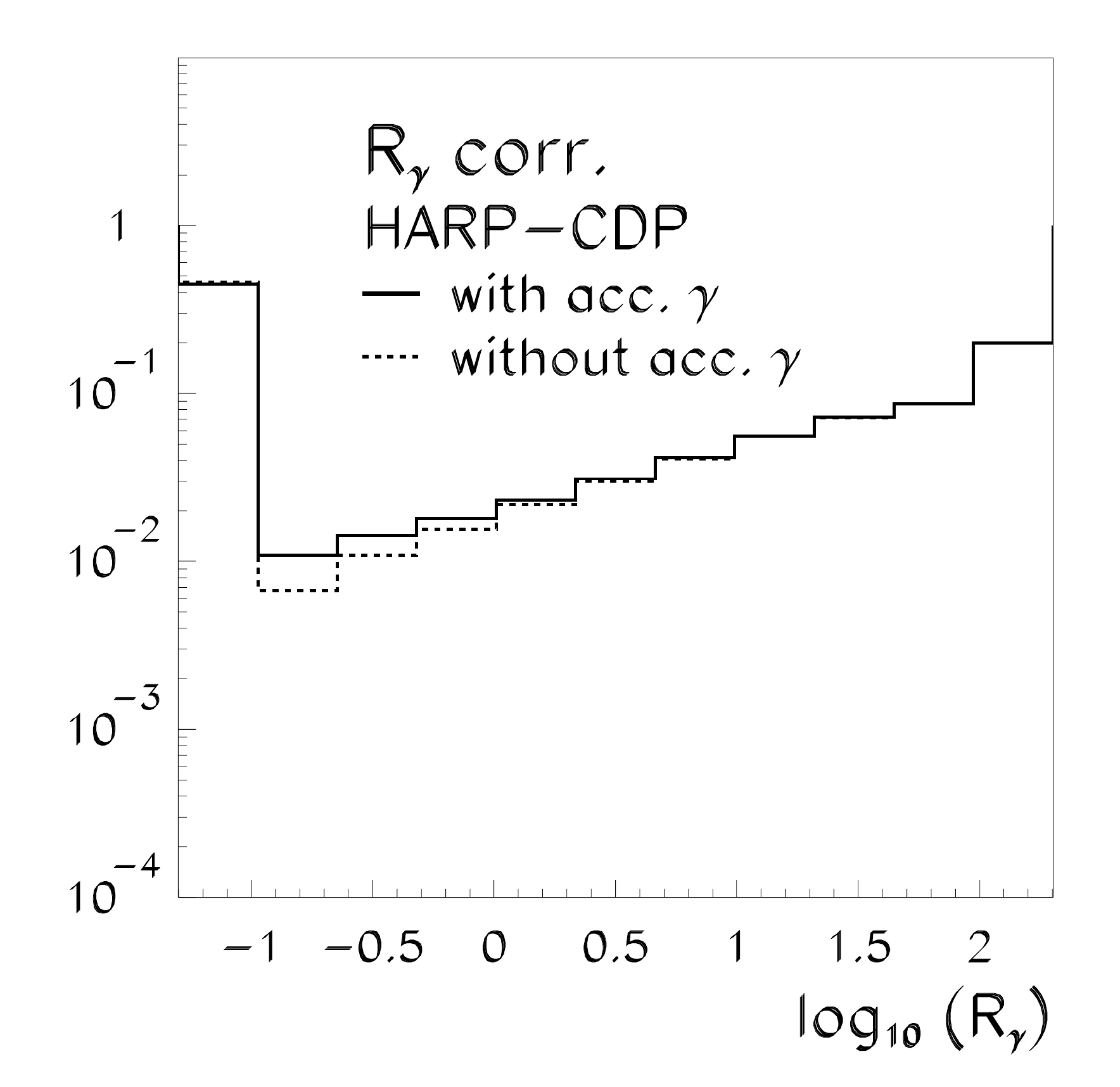} \\   
\includegraphics[width=0.6\textwidth]{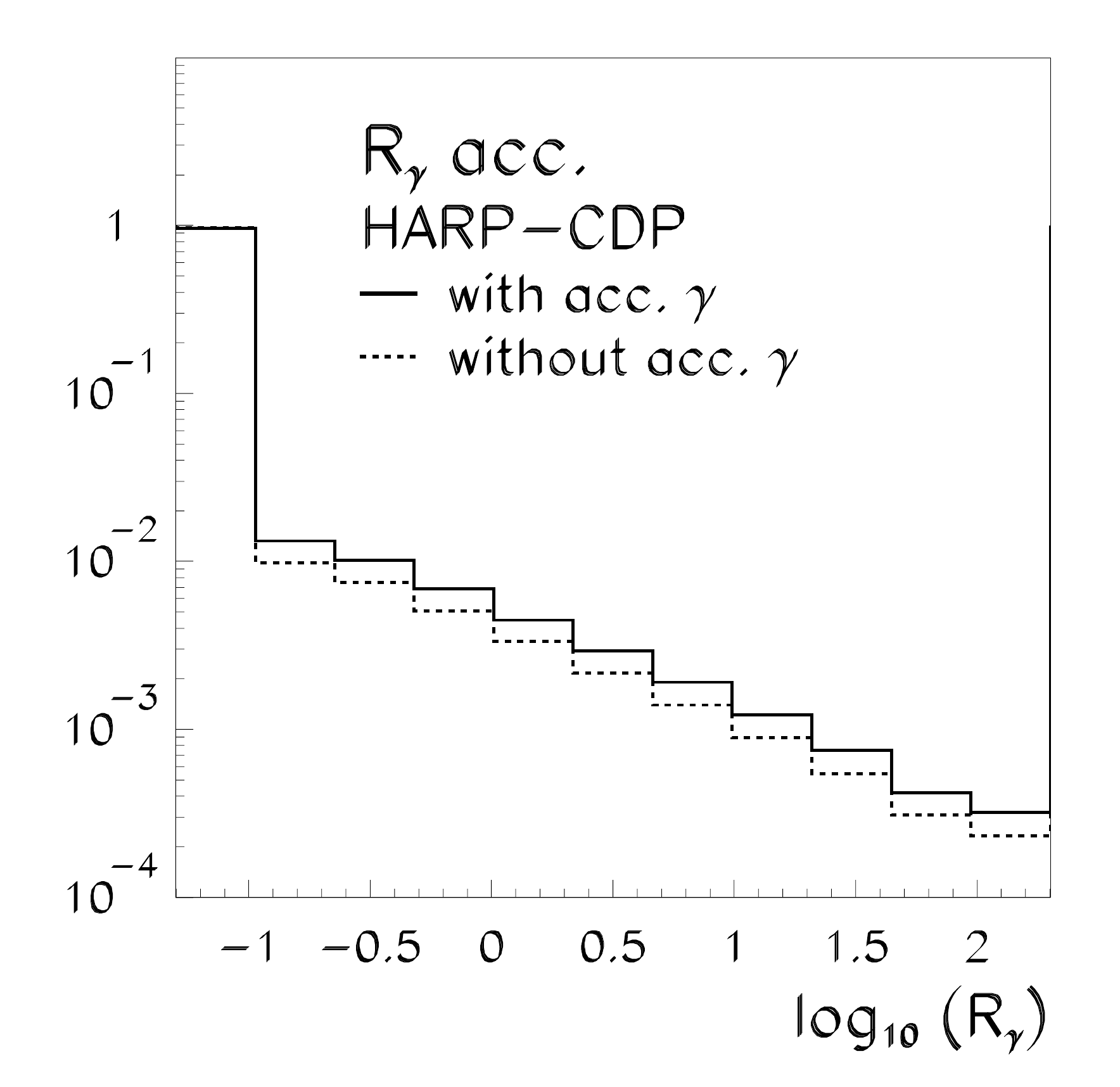} \\   
\end{tabular}  
\caption{The normalized HARP--CDP \Rgam\ hypotheses for correlated (upper panel) and accidental (lower panel) $\gamma$'s
with accidental $\gamma$'s at an effective rate of 1~kHz added (full lines), and without (broken lines).}
\label{Rgamma-wandwoaccgamma}
\end{center}
\end{figure*}

The problem of observing more than one $\gamma$ within the window 8 $\mu$s $\leq \Delta t \leq$ 1~ms
was already addressed. Because of LSND's choice that in case of more than one \gam\ being observed, the one with
the largest \Rgam\ is used, the \Rgam\ distribution is altered depending on the rate of accidental $\gamma$'s.
While LSND do not discuss this issue at all, we show in Fig.~\ref{Rgamma-wandwoaccgamma} the
HARP--CDP \Rgam\ hypotheses for correlated and accidental $\gamma$'s
with and without accidental $\gamma$'s at an effective rate of 1~kHz added. The difference is visible
and its effect on the `beam excess' will be discussed in Section~\ref{fitresults}.

\begin{figure*}
\begin{center}
\includegraphics[width=0.8\textwidth]{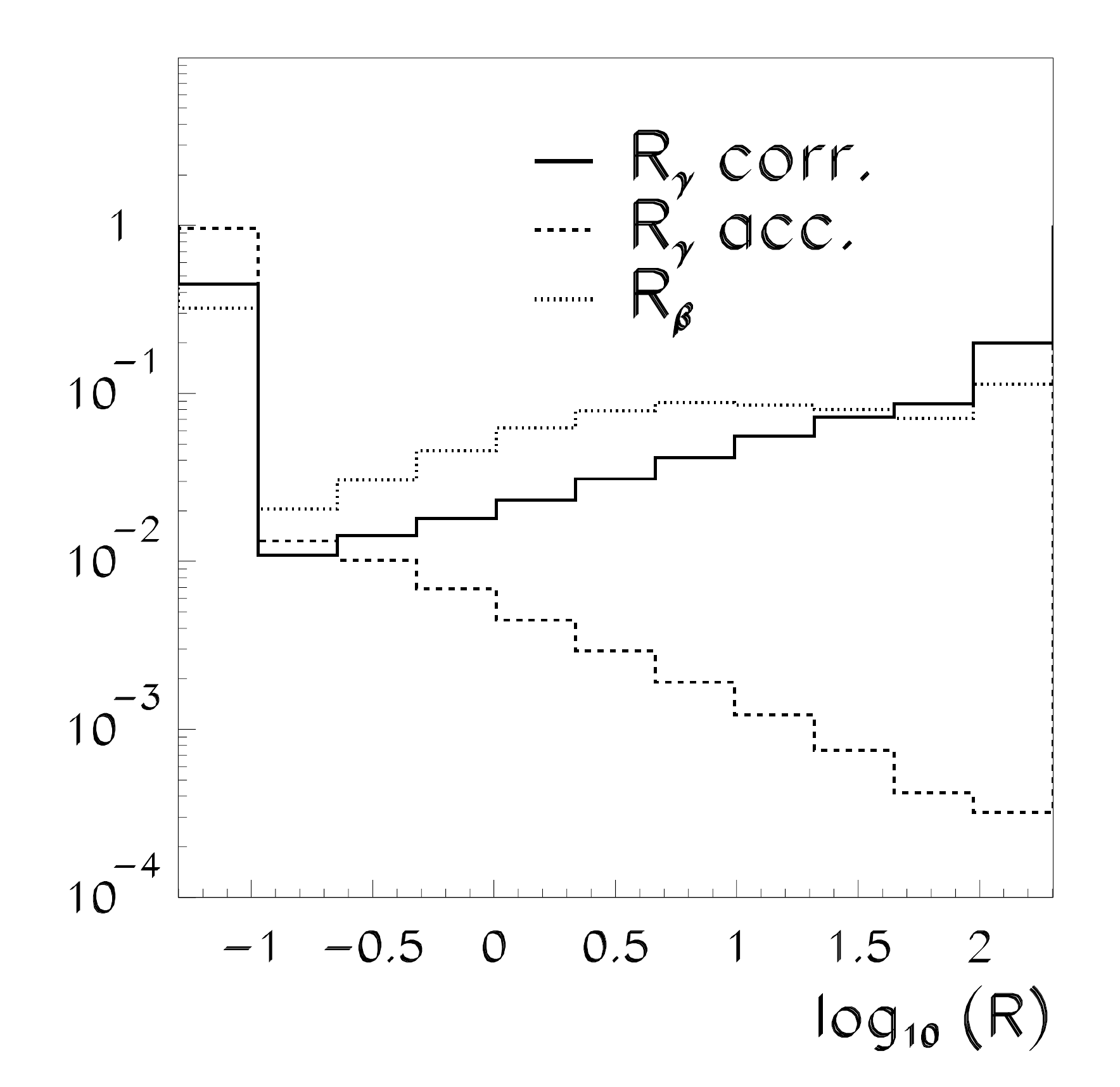}    
\caption{Comparison of the normalized HARP-CDP $R_{\gamma}$ distributions of correlated and uncorrelated
$\gamma$'s, and of the $R_{\beta}$ distribution stemming from misidentified positrons from
$^{12}$N$_{\rm gs}$ beta decays.}
\label{Rgamma-corraccbeta}
\end{center}
\end{figure*}

Next, Fig.~\ref{Rgamma-corraccbeta} shows a comparison of the normalized 
HARP-CDP $R_{\gamma}$ distributions of correlated and uncorrelated
$\gamma$'s, and of the $R_{\beta}$ distribution stemming from misidentified positrons from
$^{12}$N$_{\rm gs}$ beta decays. The latter is closer to the \Rgam\ distribution of correlated $\gamma$'s than 
the one for accidental $\gamma$'s which will have the consequence that the bulk of this background will be
misinterpreted as `beam excess',  if not duly subtracted.

\begin{figure*}
\begin{center}
\begin{tabular}{c}
\includegraphics[width=0.6\textwidth]{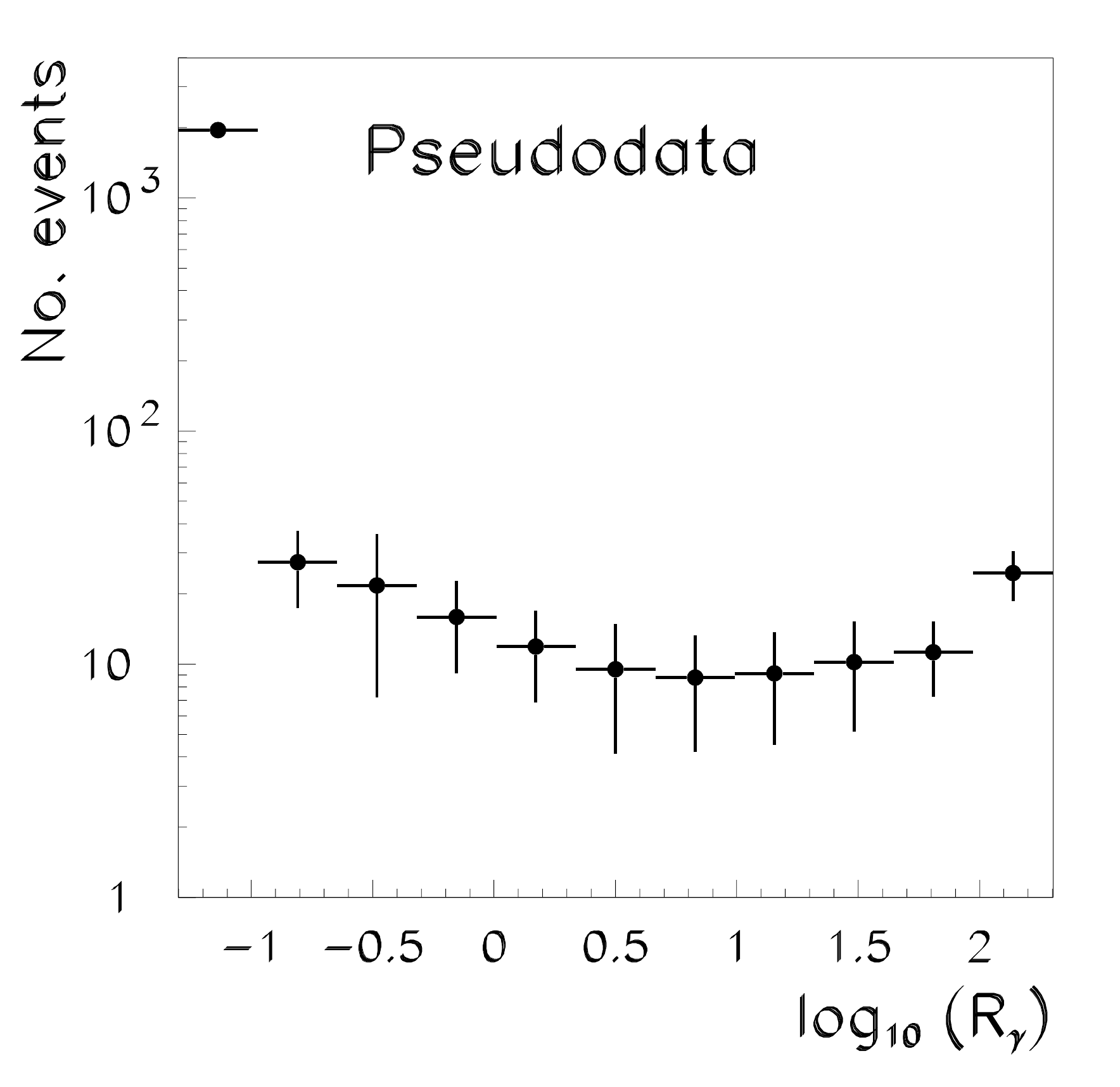} \\   
\includegraphics[width=0.6\textwidth]{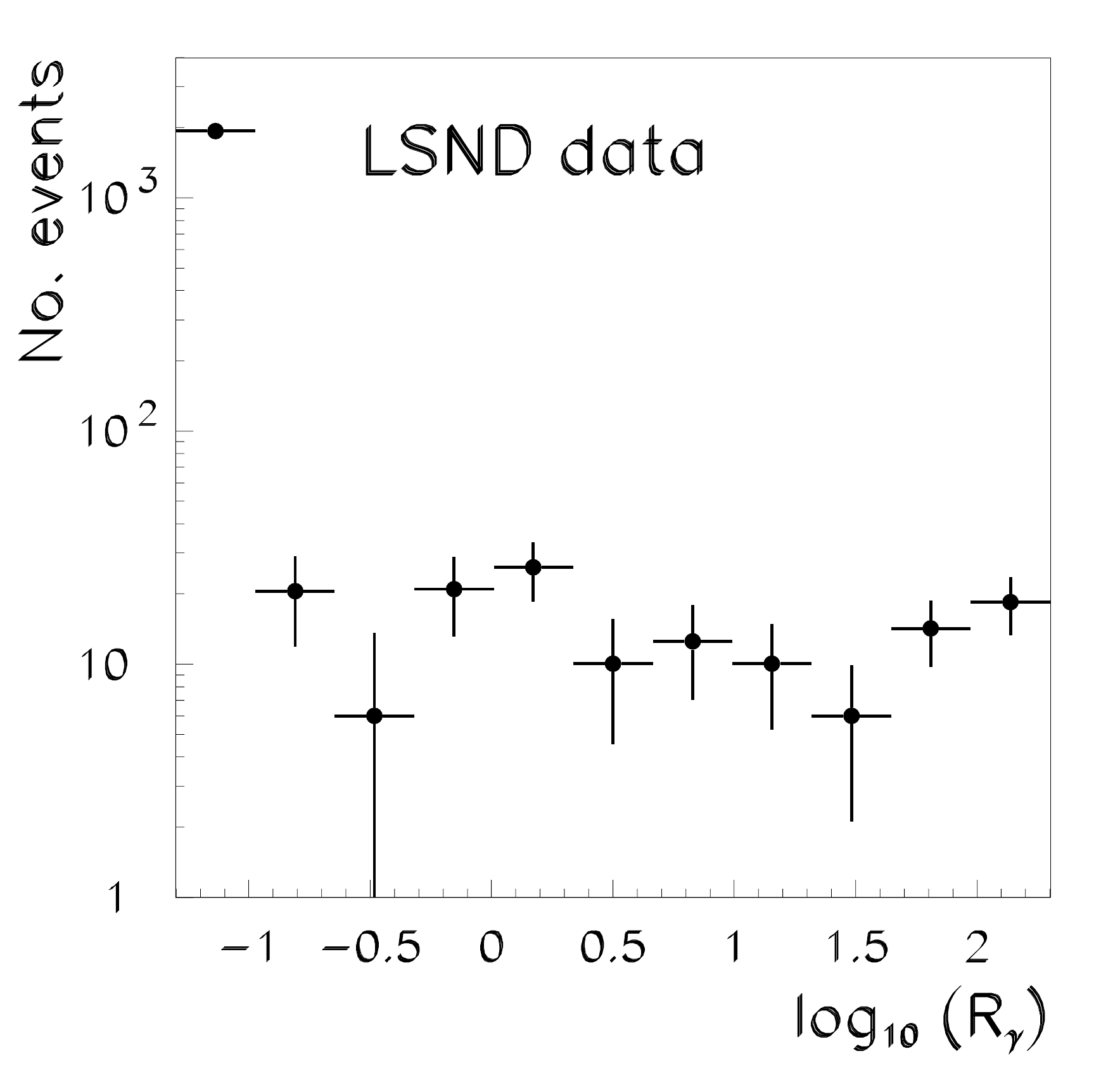} \\   
\end{tabular}  
\caption{Comparison of the `reference' $R_{\gamma}$ distribution of the HARP--CDP pseudodata (upper panel)
with the $R_{\gamma}$ distribution of the LSND data.}
\label{Rgamma-PseudoandLSNDdata}
\end{center}
\end{figure*}

Now we have all ingredients to show the `reference'
\Rgam\ distribution of the HARP--CDP pseudodata, and to compare it with the \Rgam\ distribution of the LSND data.
This comparison is shown in Fig.~\ref{Rgamma-PseudoandLSNDdata}. The LSND distribution is a graphical representation of the numbers listed
in Table~\ref{DigitizationofFig14inPRD64}.

\begin{figure*}
\begin{center}
\includegraphics[width=0.8\textwidth]{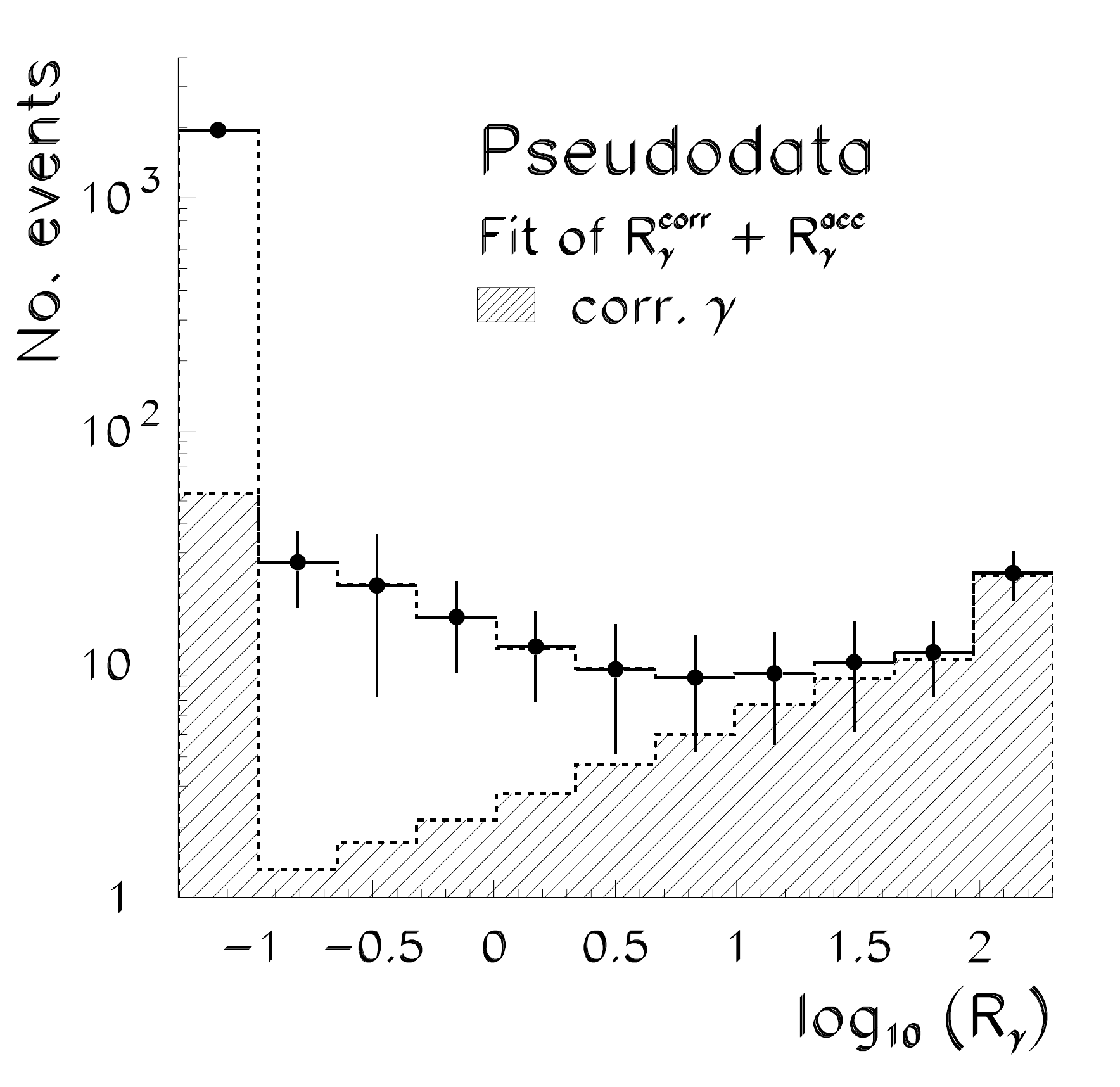}     
\caption{Fit of the pseudodata $R_{\gamma}$ distribution with the sum of the hypotheses for correlated (hatched) and accidental (open) $\gamma$'s.}
\label{Referencepseudodatafit}
\end{center}
\end{figure*}

We must demonstrate the consistency of our approach: the fit of the `reference' $R_{\gamma}$ distribution of the HARP--CDP pseudodata with the HARP-CDP \Rgam\ hypotheses must return the input 
beam excess of 120 events. This expectation is met:  the respective fit result is numerically presented in Table~\ref{Tablereferencepseudodatafit} and graphically shown in 
Fig.~\ref{Referencepseudodatafit}.
\begin{table*}[htdp]
\caption{Selected fits of pseudodata}
\begin{center}
\begin{tabular}{|c|c|c|c|}
\hline
Fit No.   &  No. of events with corr. $\gamma$ & corr. coeff. & Comment \\
\hline
\hline
5 & $120.2  \pm  21.9$   & $-0.165$ &  Fit of `reference' pseudodata \\    
5a & $120.2 \pm 22.6$            & $-0.256$ &  Fit No.~5 without 1st $R_{\gamma}$ bin \\
\hline
6 &  $126.3  \pm  22.0$  & $-0.156$ & `Base distributions' not modified by acc. $\gamma$'s \\ 
31 &   $122.5  \pm  22.1$ &  $-0.165$  & includes 2.7 positrons from $^{12}$N$_{\rm gs}$ \\
\hline
\end{tabular}
\end{center}
\label{Tablereferencepseudodatafit}
\end{table*}

Fit No.~6 employs \Rgam\ hypotheses not modified by the inclusion of accidental $\gamma$'s with an effective
rate of 1~kHz. This fit is motivated by the observation that LSND do not discuss this important modification
although in our view the inclusion is mandatory. `Base distributions', by construction, do not include accidental 
$\gamma$'s. The \Rgam\ distribution of the data and pseudodata, respectively, does. Hence the \Rgam\ hypotheses
must, too. Fit No.~6 shows that the non-inclusion of accidental $\gamma$'s would produce a larger `beam excess'
by some 6 events. However, since we have no evidence that LSND did not include properly the observed effective rate
of accidental $\gamma$'s, no further use is made of the result of Fit No.~6.

Fit No.~31 adds to the pseudodata 2.7 positrons from $^{12}$N$_{\rm gs}$ along the lines of the discussion
in Section~\ref{positrons}, with the pertinent $R_{\beta}$ distribution. The fit hypotheses comprise only correlated and accidental $\gamma$'s. Since the $R_{\beta}$ distribution is much closer
to the $R_{\gamma}$ distribution of correlated $\gamma$'s than the one of accidental $\gamma$'s, 2.3 events out of
the 2.7 events are interpreted as correlated $\gamma$'s which means that LSND's fit result for the `beam excess'
has a background of 2.3 events that was not subtracted.

\subsection{Pseudodata fit results}
\label{fitresults}

In Table~\ref{Pseudodatafitresults}, we present the results of fits of pseudodata sets that we selected
as most representative out of a large variety that we studied, and that we used to derive our error estimates. 

Fit numbers with `a' and `b' refer to a pair of fits where a parameter was varied symmetrically up and down.

The `beam excess' is given together with the correlated statistical error that reflects the input 
errors of the 11 \Rgam\ bins of the respective pseudodata.

In the case of pairs of fits, labelled `a' and `b' in Table~\ref{Pseudodatafitresults}, systematic
errors are listed only for the 
latter to avoid double counting. Systematic errors are calculated from the changes of the `beam excess' with 
respect to the best-fit value of the
`reference' pseudodata Fit No.~5, 120.2 events (see 
Table~\ref{Tablereferencepseudodatafit}). In the case of Fit No.s~32a and 32b, the systematic error is calculated
from the changes of the `beam excess' with respect to the best-fit value including the $^{12}$N$_{\rm gs}$
background, 122.5 events (see Fit No.~31 in Table~\ref{Tablereferencepseudodatafit}).

Overall errors are obtained by quadratic addition of the errors shown in Table~\ref{Pseudodatafitresults}. 
The statistical error of 21.9 events of the `reference' pseudodata Fit No.~5 leads together with 
the systematic error of 17.3 events to a total error of the `beam excess' of 27.9 events.

At the same time, a subtraction of 2.3 primary electrons with positrons from $^{12}$N$_{\rm gs}$ beta decays that are misidentified as correlated $\gamma$ from neutron capture, reduces the LSND `beam excess' from 117.9 to 115.6 events.
\begin{table*}[htdp]
\caption{Results of fits of  selected pseudodata sets}
\begin{center}
\begin{tabular}{|c|c|c|c|c|}
\hline
Fit No.   & `beam excess' & corr. coeff. & Syst. error & Comment \\
\hline
\hline  
  21a &  $128.7  \pm  22.6$ &   $-0.168$ &                       
      & corr. $\gamma$ `efficiency' up by 7\% \\
  21b &  $111.8  \pm  21.2$ &    $-0.161$ & 8.5  
      & corr. $\gamma$ `efficiency' down by 7\% \\
\hline      
  23a &  $121.8  \pm  22.0$ &    $-0.164$ &                          
      & acc. $\gamma$ `efficiency' up by 7\% \\
  23b &  $118.6  \pm  21.9$ &    $-0.166$ & 1.6  
      & acc. $\gamma$ `efficiency' down by 7\% \\
\hline
  32a &  $ 123.2  \pm  22.2$ &   $-0.165$ &                          
      & positrons up by 30\% \\
  32b &   $121.9  \pm  22.1$ &   $-0.165$ &  0.7  
      & positrons down by 30\% \\ 
\hline
  36a &   $122.5  \pm  22.1$ &   $-0.164$ &                          
      & acc. $\gamma$ rate 1.2~kHz \\
  36b &   $118.1  \pm  21.8$ &   $-0.166$ &  2.2  
      & acc. $\gamma$ rate 0.8~kHz \\
\hline
  41a   &   $134.0  \pm  23.2$ & $-0.175$   &                 
      & spatial $\gamma$  resolution $\sigma = 35$~cm \\
  41b   &   $112.7  \pm  21.1$ & $-0.157$   &  11.1               
      & spatial $\gamma$  resolution $\sigma = 45$~cm \\
\hline
  53  &     $127.9  \pm  22.4$ &   $-0.163$ & 7.6  
      & slower rise of $\Delta r$ of acc. $\gamma$'s \\
\hline
  61  &     $122.1  \pm  22.0$ & $-0.164$   & 1.9   
      & large $N_{\rm hits}$ of acc. $\gamma$'s up \\  
\hline
  72a &   $125.4  \pm  22.3$   & $-0.165$ &                           
      & small $N_{\rm hits}$ of $\gamma$'s down \\
  72b &   $113.5  \pm  21.5$ &   $-0.164$ &  6.0 
      & small $N_{\rm hits}$ of $\gamma$'s up  \\    
\hline
\hline
   &  &  &  17.3  & \\ 
\hline    
\end{tabular}
\end{center}
\label{Pseudodatafitresults}
\end{table*}

Table~\ref{OverallLSNDsummary} compares the main results of the LSND analysis with our
assessment of the situation, discussed in the preceding paper~\cite{firstpaper} and in this paper.
Our conclusion is that the significance of the `LSND anomaly' is not 3.8~$\sigma$ but 2.3~$\sigma$, 
or rather---in view of further concerns that were 
put forth at several occasions but not quantitatively followed up---not larger than 2.3~$\sigma$.
 \begin{table*}[h]
 \caption{The `LSND anomaly' and its significance}
 \label{OverallLSNDsummary}
 \begin{center}
 \begin{tabular}{|l|c|c|}
 \hline
  & LSND published & This paper's analysis \\
 \hline
 \hline
 `Beam excess' & $117.9 \pm 22.4 $ &  $115.6 \pm 27.9 $ \\
 \hline
 \hline
 Background I & $19.5 \pm 3.9 $ &  $30.6 \pm 8.8 $ \\
 \hline
 Background II & $10.5 \pm 4.6 $ &  $13.8 \pm 8.2 $ \\
 \hline
 \hline
`LSND anomaly' & $87.9 \pm 23.2 $ &  $71.2 \pm 30.4 $ \\
Significance & $3.8\:\sigma$ & $2.3\:\sigma$ \\ 
 \hline
 \end{tabular}
 \end{center}
 \end{table*}

\section{Conclusion}

A summary of our concerns about the published results from the LSND experiment reads as follows. The first
major concern is the underestimation of the standard \anue\ flux which is caused by (i) inadequate knowledge of
$\pi^-$ production by 800~MeV/{\it c} protons on various target nuclei, at the time when the LSND experiment was conducted, and (ii) from not taking into account $\pi^-$ production by neutrons. Both effects 
are quantitatively discussed in our preceding paper~\cite{firstpaper}. The second major concern is
what we consider shortcomings in the LSND data analysis, discussed in this paper: questions on the `efficiencies' of correlated and accidental $\gamma$'s; questions on the effective rate of  accidental $\gamma$'s
and their influence on the \Rgam\ hypotheses for correlated and accidental $\gamma$'s; 
missing positrons from $^{12}$N$_{\rm gs}$ beta decays that are misidentified as correlated $\gamma$'s;
and missing systematic errors of the `base distributions'. Our conclusion is that the significance of the `LSND anomaly' is not 3.8~$\sigma$ but not larger than 2.3~$\sigma$.

\section*{Acknowledgements}

We express our sincere gratitude to HARP's funding agencies 
for their support.

\end{document}